\documentclass[11pt]{article}
\usepackage{epic}
\usepackage{latexsym} \usepackage{epsf}
\usepackage{a4}
\usepackage{amsfonts}
\usepackage{epic}
\usepackage{color}
\usepackage[colorlinks]{hyperref}
\usepackage{hyperref}
\def\be{\begin{equation}}
\def\ee{\end{equation}}
\def\bea{\begin{eqnarray}}
\def\eea{\end{eqnarray}}
\def\dd{\partial}
\def\'{\prime}
\def\a{\alpha}

\def\ba{\begin{array}}
\def\ea{\end{array}}

\def\nn{\nonumber}
\def\II{\hbox{{1}\kern-.25em\hbox{l}}}
\begin{document}

{\begin{center}
\begin{title}
\bf{\LARGE{Fusion Rules of the Lowest Weight Representations of
$osp_q(1|2)$ at Roots of Unity}: {\Large{Polynomial Realization
}}}

\end{title}
\end{center}
\vspace{3mm}

{\begin{center}

{{\large D.~Karakhanyan, \footnote{e-mail:
karakhan@mail.yerphi.am}
 Sh.~Khachatryan\footnote{e-mail: shah@mail.yerphi.am}} \\
[3mm] Yerevan Physics Institute , \\
Br.~Alikhanian st.2, 375036, Yerevan, Armenia.}
\end{center}
}

\vspace{0.5cm}
\begin{abstract}

 The degeneracy of the lowest weight representations of the quantum
superalgebra $osp_q(1|2)$ and their tensor products at exceptional
values of 
$q$
is studied. The main features of the structures of the finite
dimensional lowest weight representations and their fusion rules
are illustrated using realization of group generators as
finite-difference operators acting in the space of the
polynomials. The complete fusion rules for the decompositions of
the tensor products at roots of unity are presented. The
appearance of indecomposable representations in the fusions is
described using Clebsh-Gordan coefficients derived for general
values of $q$ and at roots of unity.
\end{abstract}



\section{Introduction}

The quantum algebras are studied intensively starting with that
very moment they were invented by Faddeev and Takhtajan and et al
in 1981 \cite{FT}. Since that time the quantum algebras found
numerous applications in different fields of physics and
mathematics and are related by thousands links with other branches
of science.

Being special, quantum algebras (or superalgebras) have in many
cases the same representations as corresponding classical
(non-deformed) Lie algebras, but along with that, new, quite
different representations appear in quantum case [2-23]. In
standard deformation scheme (with dimensionless deformation
parameter $q$) the center of the algebra is enlarged and  new
Casimir operators and correspondingly  new type of representations
appear when $q$ is given by a root of unity
\cite{PS,S,GL,PP,Ar1,Ge,Zh,A,L,KKH,RRT,JimGRAS,chakap}. However
the structure of  the representation space is not only thing,
which is subjected to deformation, the decomposition of the tensor
product of the representations is deformed too.  The allowed
values of "spin" of the $q$-deformed finite-dimensional
representations are restricted when deformation parameter $q$ is
given by a root of unity. In this case a proper subspace can
appear inside of representation $V$, irreducible at general $q$,
making the latter non-irreducible. As a consequence some items in
the decomposition of the tensor products are unified into new,
indecomposable representations $\mathcal{I}$ \cite{PS, S, Ar1,
KKH}.
 Although there is a considerable amount of work (partly
cited above) devoted to the representation theory  at the roots of
unity, it seems a thorough investigation of the fusion rules
regarding to the all possible representations is needed
(especially a detailed analysis of $V\otimes \mathcal{I}$ and
$\mathcal{I} \otimes \mathcal{I}$).

In this article we clarify the mentioned aspects in visual form
for the lowest weight representations of the quantum superalgebra
$osp_q(1|2)$ \cite{KR,S,Ge,Zh,A,Ki,SHKKH,z,SB}. The 
 orthosymplectic superalgebras $osp(n|2m)$ (and their quantum
deformations) are actual
 in CFT, in the theory
of integrable models, in the string theory (see the works
\cite{S,KMW} and references therein). The change of the
representation's spectrum at roots of unity brings to new
peculiarities, for instance to new solutions of the Yang-Baxter
equations in the theory of $2d$ integrable models
\cite{JimGRAS,chakap,SHKKH}.

There is a correspondence between the irreps of the graded algebra
$osp_q(1|2)$
 and
 the irreps of
 the algebra $sl_{i q^{1/2}}(2)$
 \cite{S,Ki,z,SHKKH} at general $q$, which allows to use
some techniques evolved from the study of the quantum deformation
of $sl(2)$ \cite{PS,S,KKH,KK} for investigation of the
super-algebra $osp_q(1|2)$.
 And our approach
consists in studying the finite-dimensional representations and
their tensor products at general values of $q$, to find out at the
complex $q$-plane the "singular points" (located on the unit
circle) of these representations or their tensor products'
decompositions.

 The finite-difference realization of the group
generators, acting on the space of the polynomials, provides clear
and compact description and answers the purposes in the best way.
The approach of the projection operators gives 
a simple understanding why degeneracies appear in the
decompositions of the representations' tensor products. Being
comprehensible enough constructions presented here are adapted
especially for the physical applications.

The paper is organized as follows: in the second section the
algebra generators and the co-product in the mentioned polynomial
realization are presented and in the third section some 
examples of the representations and their tensor products are
considered in details. The discussed patterns illustrate the
principal cases, including multiple tensor products of the irreps
and indecomposable representations. In the next section an 
analysis of the fusion rules from the viewpoint of the projector
operators is performed. The last sections are devoted to general
analysis and conclusions based on the observations of the previous
sections, accompanied with explicit proof of the general formulas.
In the subsections $5.1$, $5.2$ we represent indecomposable
representations emerging from the tensor product's decompositions 
of the odd 
dimensional irreps, give detailed computation of their dimensions,
state the crucial principles how they appear in the fusions, and
propose fusion rules for decompositions. In the subsection $5.3$
Clebsh-Gordan coefficients in general form are derived, by means
of which the fusion rules' degenerations can be presented by
direct
constructions 
(see Appendix).
  The
 section $6$ 
 is devoted
 to the even dimensional representations and to the
 correspondence
between the representations of the quantum deformations of the
algebras $osp(1|2)$ and $sl(2)$
at the
exceptional values of the deformation parameter%
.

\section{The algebra $osp_q(1|2)$ and co-product: polynomial realization}

The quantum algebra $osp_q(1|2)$ \cite{KR,S,Ki} is a Hopf algebra.
It is generated by two odd generators $e$, $f$ and the even
generators $k,\; k^{-1}$, which obey the following
(anti-)commutation relations:
 \be\label{alg} f k^{\pm 1}=q ^{\pm 1}k^{\pm 1}f,\quad
 e k^{\pm 1}=q ^{\mp 1}k^{\pm 1}e, \quad
\{e,f\}=\frac{k-k^{-1}}{q-q^{-1}}=[H]_q,
 \ee
where $q\in\mathbb{C}$, $q\neq0,\pm1$, $[a]_q=\frac{q^a-q^{-a}}
{q-q^{-1}}$, and we use the notation $k^{ \pm1}=q^{\pm H}$ to keep
connection with the non-deformed case. The bracket 
$\{,\}$ stands for 
anti-commutator. When $q\to 1$ these relations reduce to the
ordinary (anti-)commutation relations of the super-algebra
$osp(1|2)$ for the simple
generators $e, f, H$.
%

The quadratic Casimir operator $c$, given by the formula
 \be\label{cas}c=-(q+2+q^{-1})e^2f^2+
(k q^{-1}+q k^{-1})ef+[H-\frac{1}{2}]_q^2=\left(
(q^{\frac12}+q^{-\frac12})ef-[H-\frac12]_q\right)^2,
 \ee
  is the square of a simpler operator, the so called
Scasimir operator \cite{L,A}.

The algebra generators can be represented as finite-difference
operators on the graded space of the polynomials:
 \be\label{difop}
k^{\pm 1}=q^{\pm(2x\dd+\theta\dd_\theta-2j)}=q^{\pm(2x\dd-2j)}+q^{
\pm(2x\dd-2j)}(q^{\pm1}-1)\theta\dd_\theta,\qquad f=\dd_\theta+
\frac\theta x[x\dd]_q,
 \ee
$$
e=(x\dd _\theta+\theta[ x\dd]_q)([x\dd\!+\!1\!-\!2j]_q
\!-\![x\dd\!-\!2j]_q) -[2j]_q\theta,
$$
where $\theta$ is a Grassmann variable, while $x\in\mathbb{C}$.
Note that $F=f^2=D_q=\frac 1x\frac{q^{x\dd}-q^{-x\dd}}{q-q^{-1}}$.

We fix the generator $f$ to be a lowering operator and $e$ to be a
raising operator throughout the paper, as usual. In the polynomial
representation there always exists lowest weight vector, which is
given by a constant function, i.e. the present method is
especially convenient to study the lowest weight representations.
For general values of 
$q$ odd dimensional
lowest weight representations are in one-to-one correspondence
with the representations of the non-deformed algebra $osp(1|2)$
\cite{KR,S,Ge,Ki} and  can be classified in the same way. 
 Spin-$j$ ($j\in \frac{1}{2}\;\mathbb{Z}_{+}$)
  representation with the eigenvalue $[2j+1/2]_q^2$ of Casimir operator $c$
 has dimension $4j+1$. Degeneracy occurs and new
features appear when $q$ is given by a root of unity. The center
of the algebra becomes larger. From the relations
 \bea\label{grcom}
\left[f,e^n\right.\left.\right\}&=&\left\{\ba{cc}e^{n-1}\left
[H+\frac{n-1}2\right ]_q\frac{[n]_ q[1/2]_q}{[n/2]_q}& n \mbox{ is odd}\\
e^{n-1}\left(\left [H+\frac{n}{2}\right ]_q\!-\!\left
[H+\frac{n}{2}-1\right ]_q\right)\left[\frac{n}{2}\right]_ q& n
\mbox{ is even}
\ea\right.\\\label{grcfne}\left[f^n,e\right.\left.\right\}&=&\left\{\ba{cc}f^{n-1}
\left
[H+\frac{1-n}2\right ]_q\frac{[n]_q[1/2]_q} {[n/2]_q}& n \mbox{ is odd}\\
f^{n-1} \left(\left [H-\frac{n}{2}\!+\!1\right ]_q\!-\!\left
[H-\frac{n}{2}\right ]_q\right)\left[\frac{n}{2}\right]_ q& n
\mbox{ is even}\ea\right.
 \eea
it follows that if $q^N=1$, operators $e^{\mathcal{N}}$,
$f^{\mathcal{N}}$ and $k^{\mathcal{N}}$ commute with algebra
generators, where $\mathcal{N}=2N$ for odd $N$ and $\mathcal{N}=N$
for even $N$. Here $\left[a,b\right. \left.\right\}$ is a graded
commutator, which is an anti-commutator, when both $a$ and $b$ are
odd operators, and is a commutator otherwise:
$$\left[a,b\right.\left.\right\}=a b-(-1)^{p(a)p(b)}b a.$$
 $p(a)$ is
the parity of the homogeneous element $a$ of the graded algebra,
and equals to $0$ for even (bosonic) elements and equals to $1$
for odd (fermionic) elements.

The $osp(1|2)_q$ as a Hopf super-algebra possesses
  co-unit $\epsilon$ and antipode $\gamma$ \cite{Ki}
defined as
\bea \label{counit}
&\varepsilon(e)=0,\quad\varepsilon(f)=0,\quad\varepsilon(k)=1,\quad\varepsilon(k^{-1})=1,&\\\nn
&\gamma(e)=- e k,\quad\gamma(f)=- k^{-1} f,
\quad\gamma(k)=k^{-1},\quad \gamma(k^{-1})=k. &\eea

The co-associative co-product compatible with (\ref{counit}) is
given as follows
\bea \label{coprod1} \Delta(k)=k\bar{\otimes} k,\qquad
\Delta(k^{-1})= k^{-1}\bar{\otimes} k^{-1}\qquad
\Delta(f)=f\bar{\otimes}1 +k\bar{\otimes} f, \qquad \Delta(e)=e
\bar{\otimes} k^{-1}+1\bar{\otimes} e.
 \eea
Here $\bar{\otimes}$ denotes super-tensor product for the graded
operators. On the product of two graded spaces of the polynomials
$V^i$, $i=1,2$, generated by variables $x_i, \theta_i$, the graded
nature of the super-algebra is taken into account automatically
via the inclusion of the Grassmann variables (see (\ref{difop})):
 \be\label{coprod}
k=k_1k_2,\qquad k^{-1}=k_1^{-1}k_2^{-1}\qquad f=f_1+k_1f_2, \qquad
e=e_1k_2^{-1}+e_2,
 \ee
where $\{k_i, k_i^{-1} e_i, f_i\}$ are the generators
(\ref{difop}) acting on the spaces $V_i$ correspondingly.

For the homogeneous elements $a_i,\;c_i$ of the algebra the
multiplication law for the graded tensor products \cite{ks} is
 \bea
(a_1\bar{\otimes}c_1)(a_2\bar{\otimes}c_2)=(-1)^{p(c_1)p(a_2)}(a_1
a_2\bar{\otimes}c_1 c_2).
 \eea
From the above relation and the co-product (\ref{coprod1}) the
following equations can be derived
 \be \label{fN}\Delta(f^{n})=\sum_{r=0}^{n}
[^{n}_{r}]_{-q^{-1}}(f\bar{\otimes} 1 )^{{n}-r}(k\bar{\otimes}
f)^r,\qquad \Delta(e^{n})=\sum_{r=0}^{n}
[^{n}_{r}]_{-q^{-1}}(1\bar{\otimes} e
)^{{n}-r}(e\bar{\otimes}k^{-1})^r.\ee
 Here $[^{n}_{r}]_{-q^{-1}}$ are
$q$-binomial coefficients \cite{KS}
\be\label{[]}[^{n}_{r}]_{q}=\frac{[n]_{q^{1/2}}!\;q^{(n-r)r/2}}{[r]_{q^{1/2}}!\;
[n-r]_{q^{1/2}}!},\ee
with $q$-factorials $[p]_q!=[p]_q[p-1]_q...[1]_q$.

 If the super-algebra elements
$a,\;c$ have the matrix representations $a_i^j ,\;c_i^j$ in the
representation spaces $V$ and $U$ respectively, which have basis
states $v_i$ and $u_i$, then the matrix representation of
$(a\bar{\otimes}c)$ in the representation space $V\bar{\otimes} U$
with basis states {$v_i\bar{\otimes} u_j$ is
  \bea\label{gradtp}
(a\bar{\otimes}c)_{ij}^{kr}=a_i^k c_j^r(-1)^{p(k)(p(j)+p(r))},
\quad p(i')=0,1;\quad i'=i,j,k,r.
 \eea
Here $p(i')$ is the parity of the $i'$-th basis element.
In the later discussion for simplicity we shall use usual notation
$\otimes$ for the tensor product of the graded representations.

To establish the correspondence between the polynomial realization
and the matrix formulation it is enough to assign the following
columns to the vectors of the fundamental multiplet
$\{1,\;\theta,\;x\}$: $ 1={\left(0,0,1\right)}^\tau,\quad\theta
={\left(0,1,0 \right)}^\tau, \quad x={\left(1,0,0\right)}^\tau $
 ($\tau$ stands for transposition operation).

We are going to study the lowest wight representations, arising
from the tensor products of the fundamental representations, and
for them, in case of $q$ being a root of unity, $e$ and $f$ are
$\mathcal{N}$-nilpotent: $e^{\mathcal{N}}=0$, $f^{\mathcal{N}}=0$
(it follows from (\ref{grcom}), (\ref{coprod1}, (\ref{fN})) and
from the existence of the lowest and highest weight vectors), and
$k^{\mathcal{N}}=1$. Note, that in the case when $q^N=-1$ and $N$
is odd number, it is evident from (\ref{grcom}), that the
operators $e^N,\;f^N,\;k^{\pm N}$, similar to Scasimir operator,
anti-commute with the part of the algebra generators, and commute
with the other part.
  However, in the mentioned representation
spaces
, being of interest to us, the operators $e^N,\;f^N$ ($q^N=-1$,
$N$ is odd) also can be regarded as Casimir operators with $0$
eigenvalues (here $k^N=\pm 1$).

\paragraph{The lowest weight representations at general $q$.}

One  can see that the number of bosonic states in
$(2n+1)$-dimensional irrep
$\{1,\;\theta,\;x,...,x^{n-1}\theta,\;x^n\}$, $n\in\mathbb{N}$,
exceeds the number of fermionic states by one (as the lowest
weight vector $1$ is bosonic). The quantum algebra $osp_q(1|2)$
possesses the "supersymmetric" even-dimensional representations,
$\{1,\;\theta,\;x,..., x^{n-1}\theta,\;x^n,\;x^n\theta\}$, with
equal number of fermionic and bosonic basis states as well.
Indeed, here it is necessary to have the fermionic vector
$x^n\theta$ at some $n\in\mathbb{N}$ as a highest weight vector
and one demands:
 \be\label{edrc}
e\cdot x^n\theta=0,\qquad\qquad {\rm{or}}\qquad\qquad
[2j-n]_q-[2j-n-1]_q=0.
 \ee
Equation (\ref{edrc}) has solutions, when  $q^{4j}=-q^{2n+1}$. In
other words $r=2n+2$-dimensional irreducible representations form
a sequence labelled by positive integer $r$ or by $j_r$ (see
\cite{Ki}):
 \be\label{lambda}
2j_r=n+\frac12+\lambda=\frac{r-1}{2}+\lambda,\qquad\qquad
q^\lambda=i,\qquad\qquad \lambda=\frac{i\pi}{2\log q}.
 \ee
So one sees that this series of representations has no classic
counterpart.

The action of the algebra elements  (\ref{difop}) on the states
$x^{p},\;x^{p}\theta\;$ of the spin-$j$ irrep reads as
\bea \nn && e\cdot x^{p}=\left(\frac{ }{
}[p\;]_q([p\!+\!1-\!2j]_q-[p\!-\!2j]_q) -[2j]_q\right) x^p\theta,
\\\nn && e\cdot x^{p}\theta =\left(\frac{ }{
}[p+1-2j]_q-[p-2j]_q\right)
x^{p+1},\\
&& f\cdot x^{p}=[p\;]_q x^{p-1}\theta,\qquad\qquad\qquad f\cdot x^{p}\theta=x^{p}\\
\nn && k^{\pm 1}\cdot x^{p}=q^{\pm (2p-2j)} x^{p},\quad\quad\quad
k^{\pm 1}\cdot x^{p}\theta = q^{\pm (2p-2j+1)}x^{p}\theta.\eea
  The decomposition rule of
the tensor product of two irreps with dimensions $r_1$ and $r_2$
is obtained in the same way as for the non-deformed algebra, and
can be proved by straight construction \cite{Ki} (see the Section
$4$),
\bea V_{r_1}\otimes V_{r_2}=\bigoplus_{r=|r_1 -r_2|+1}^{r_1+r_2-1}
V_{r},\qquad \Delta r=2.\label{t}\eea

The even-dimensional representations were described for the first
time in the work \cite{Ki}, and as it is stated ibidem, the
representations can be defined up to the sign of the power index
of the eigenvalues of the generator $k$, due to an automorphism of
the algebra ($e \to -e,\; f\to f,\; k\to 1/k$).
\section{The fusion rules of the low-dimensional representations}
In this section we shall consider some simple examples to
illustrate the main phenomena: the Clebsh-Gordan decomposition of
 tensor products depends on the deformation parameter and when it
takes exceptional values the direct sum decomposition  turns into
semi-direct one in certain cases. Then the block-diagonal action
of the algebra generators on the tensor product becomes
block-triangular one. Let us start with the simplest case.
\paragraph{
$\textbf{V}_2\otimes \textbf{V}_2$.}
The tensor product of two such irreps,
$\{1,\;\theta_1\}\otimes\{1,\theta_2\}=\{1,\;\theta_2,\;\theta_1,\;\theta_1
\theta_2\}$, is decomposed into the direct sum of the states
$\{1,\;\theta_1-iq^{-\frac12}\theta_2,\;\theta_1\theta_2\}$ and
$\{\theta_1-iq^{\frac12}\theta_2\}$, corresponding to the spin
one-half and spin zero representations. It means that the
representation of Casimir operator on $V_2\otimes V_2$ is
presented as a decomposition
  \be\label{c22}
c_{2\times2}=[\frac32]_q^2P_3+[\frac12]_q^2P_1=[\frac32]_q^2\II-
([\frac32]_q^2-[\frac12]_q^2)P_1,
  \ee
over projection operators $P_i$, $P_iP_j=\delta_{ij}P_i$, defined
on the states with spin zero and spin one-half. The eigenvalues of
Casimir operator (\ref{c22}) become degenerate, when $q^2=-1$,
i.e. $[\frac32]_q^2=[\frac12]_q^2$. This degeneracy of the
eigenvalues is not accompanied by degeneracy of the eigenvectors:
vectors $\theta_1-iq^{-\frac12}\theta_2$ and
$\theta_1-iq^{\frac12}\theta_2$ still remain linearly independent.
So the rule (\ref{c22}) for the fusion of the couple of
two-dimensional representations is valid for the exceptional
values of $q$ as well.
\paragraph{
$\textbf{V}_2\otimes \textbf{V}_3$.} In this case from (\ref{t})
one obtains, that on the space of the tensor
product 
{$\{1,\;\theta_1\}\otimes\{1,\;\theta_2,\;x_2\}=\{1,\;\theta_2,\;x_2,
\;\theta_1, \;\theta_1 \theta_2,\;\theta_1 x_2\}$} the quadratic
Casimir operator can be written as a sum of the projection
operators
  \be\label{c2x3}
c_{2\times3}\!=\![2\!+\lambda]_q^{2}P_4\!+\![1\!+\lambda]_q^{2}P_2=[2\!+
\!\lambda]_q^{2}\II\!-\!([2\!+\!\lambda]_q^2-\![1\!+\!\lambda]_q^{2})P_2\!\!.,
  \ee
on the two- and four-dimensional states,
  {\bea\label{c23basis}
V_2=\{\theta_1\!-\!iq^{\frac12}\theta_2,\; (q\!-\! 1)x_2\!-\!i
q^{-\frac12}\theta_1\theta_2\},\quad V_4=\{1,\;q\theta_1 -i(q^
{\frac12}-q^{-\frac12})\theta_2,\;x_2 +iq^{\frac12}\theta_1
\theta_2,\;\theta_1x_2\}.\eea}
The projectors $P_4$ and $P_2$ have poles  at $q^3=-1$, which
means that the separation of the space spanned by vectors
$\{1,\;\theta_1,\;\theta_2,\;\theta_1\theta_2 ,\;x_2,\;
\theta_1x_2\}$ into $V_4$ and $V_2$ has no longer sense. Both the
eigenvalues and eigenvectors of the Casimir operator have
degeneracy at that point (i.e.
$[2\!+\!\lambda]_q=-[1\!+\!\lambda]_q$ and two vectors
$\{q\theta_1-i(q^{\frac12}-q^{-\frac12})\theta_2$, $x_2
+iq^{\frac12}\theta_1\theta_2\}$ belonging to the four-dimensional
representation are linearly dependent with the vectors of
two-dimensional representation space
$\{\theta_1\!-\!iq^{\frac12}\theta_2,\;(q\!-\!
1)x_2\!-\!iq^{-\frac12}\theta_1\theta_2\}$). The whole
representation space contains two more vectors linearly
independent with $V\equiv\{v_i\}=\{1,\;
\theta_1\!-\!iq^{\frac12}\theta_2,\;x_2\!+\!i
q^{\frac12}\theta_1\theta_2,\;\theta_1x_2\}_{q^3=-1}$ vectors,
 and they can be chosen as $
U=\{u_2,\;u_3\}=\{\theta_1,\;\theta_1\theta_2\}$. The action of
the generators on the whole space is
\bea\nn e\cdot{\{v_1,\;v_2,\;v_3,\;v_4,
\;u_2,\;u_3\}}&=&{\{-iq^{\frac12}v_2,
\;-iq^{\frac12}v_3,\;0,\;0,\;u_3,\;-v_4\}},\\\label{23}
f\cdot{\{v_1,\;v_2,\;v_3,\;v_4, \;u_2,\;u_3\}}&=&{\{0,\;0,\;
v_2,\;v_3,\;v_1,\;-iq^{-\frac12}v_2-iq^{-\frac32}u_2\}},\\\nn
k\cdot{\{v_1,\;v_2,\;v_3,\;v_4,
\;u_2,\;u_3\}}&=&{\{-iq^{-\frac32}v_1,
\;-iq^{-\frac12}v_2,\;-iq^{\frac12}v_3,\;-iq^{\frac32}v_4,\;-iq^{-\frac12}u_2,
\; -iq^{\frac12}u_3\}}.\eea
 In this way one sees
that at $q^3=-1$ the action of the algebra generators $G$ on the
tensor product $V_2\otimes V_3$ acquires block-triangular form:
$G\cdot V\Rightarrow V$, $G\cdot U\Rightarrow U+V$ (algebra
generators $G$ map the vectors belonging to $V$ into themselves
and map the vectors forming $U$ into the vectors of the spaces $U$
and $V$). So $V_2\otimes V_3$ at $q^3=-1$ has to be considered
itself as an indecomposable six-dimensional representation, with
proper sub-representation $V_4$: we denote it by
$\widehat{\bar{V}_4\oplus V_2}$ or by
${\mathcal{I}}^{(6)}_{\{4,2\}}$. Here the "bar" over $V_4$ means
that $\{v_1,\;v_2,\;v_3, \;v_4\}$ at $q^3=-1$ is not irreducible
and contains invariant two-dimensional subspace $\{v_2,\;v_3\}$
(see (\ref{23})). ${\mathcal{I}}^{(6)}_{\{4,2\}}$ has two lowest
and two highest weights. So, we find
  \be
V_2\otimes V_3=\left\{ \ba{ll}\mathcal{I}^{(6)}_{\{4,2\}},&\mbox
{if}\quad q^3=-1,\\V_2\oplus V_4,& \mbox{for other
cases},\ea\right.
  \ee
Note that Casimir operator (\ref{c2x3}) remains regular at
$q^3=-1$, but it is not longer diagonal on $\{V,U\}$:
$$
(c_{2\times3}+\frac13\II)\cdot{\{v_1,\;v_2,\;v_3,\;v_4,
\;u_2,\;u_3\}}= {\{0, \;0,\;0,\;0,\;2q v_2,\;
2iq^{\frac32}v_3\}}|_{q^3=-1},
$$
acquiring triangular form on the vectors with weights
$h=\pm\frac12+\lambda$.

{\textit{Definition}.} Hereafter, by
${\mathcal{I}}^{(r)}_{\{k,r-k\}}$ (letting $k > r-k$) we shall
denote the non irreducible representations of dimension $r$, which
appear in the fusions instead of the direct sum $V_k\oplus
V_{r-k}$, when $q$ takes exceptional values (see for detailed and
general descriptions the last section).

\paragraph{ 
$\textbf{V}_3\otimes \textbf{V}_3$.}

Acting by finite-difference operators (\ref{cas}), (\ref{difop})
on the tensor product of two spin one-half representations
$\{1,\theta_1,x_1\}$ and $\{1,\theta_2,x_2\}$, one can calculate
the eigenvectors 
of the Casimir operator 
 \bea\label{v35}
V_5=\{{\varphi_5}^\a\}\equiv\{1,\theta_1+q\theta_2,x_1+(q-q^2)\theta_1
\theta_2+q^2x_2,x_1\theta_2+qx_2\theta_1,x_1x_2\},\qquad
\quad\\\nn V_3=\{{\varphi_3}^i\}\equiv\{q\theta_1-\theta_2,x_1+
(1+q)\theta_1\theta_2-x_2,qx_1\theta_2-x_2\theta_1\},\;\;\:
{V_1}=\{x_1+\theta_1\theta_2-q^{-1}x_2\}. \eea
 Here $\varphi$'s
are the eigenvectors of $c$, and $\a=1,2,\ldots,5,\quad i=1,2,3$.
In this way one explicitly constructs $V_3\otimes V_3$ tensor
product decomposition as the sum of representations $V_1\oplus
V_3\oplus V_5$
 \be
c_{3\times3}=[\frac52]^2_q P_5+[\frac32]^2_qP_3+ [\frac12]^2_q
P_1=[\frac52]^2_q I+([\frac32]^2_q-[\frac52]_q)
P_3+([\frac12]^2_q-[\frac52]^2_q)P_1.\label{c33}
 \ee
The projection operators $P_1,\;P_3,\;P_5$ have the following
multipliers correspondingly: $ (q-1+q^{-1})^{-1}$, $
(q+q^{-1})^{-1}$ and $ (q+q^{-1})^{-1}(q-1+q^{-1})^{-1}$.
%
{{The poles of the operators $P_r$ correspond to three
different cases $q^4=1$, $q^6=1$ and $q^8=1$.
}}

When $q=\pm i$, all eigenvalues coincide each to other:
$[\frac52]^2_q=[\frac32]^2_q=[\frac12]^2_q$, and six eigenvectors
coincide each to other pairwise (degeneracy of the eigenvectors
 shows itself as a linear dependence of vectors, we
denote this relation between vectors as $\approx$):
$\{{\varphi_5}^{\alpha+1}\}\approx \{{\varphi_3}^\alpha\}$,
$\alpha=1,2,3$; two of three projectors, $P_5$ and $P_3$, become
singular, correspondingly the sum of the representations $V_5$ and
$V_3$ transforms into one new indecomposable representation,
${\mathcal{I}}^{(8)}_{\{5,3\}} =\widehat{\bar V_5\oplus V_3}$. The
set of the vectors (\ref{v35}) has to be completed by three new
vectors to form basis, which can be taken as: $
\{1,\;\theta_1+i\theta_2,\;\theta_1- i\theta_2,\;x_1+(1+i)
\theta_1\theta_2-x_2,\;x_1+\theta_1 \theta_2+ix_2,
2x_1,\;x_1\theta_2+ix_2\theta_1,\;x_1\theta_2-
ix_2\theta_1,\;x_1x_2\}$.
Consider next the case $q^3=-1$, when $[1/2]_q^2=1/3 =[5/2]_q^2$.
The projectors $P_5$ and $P_1$ are ill-defined at that point and
one can see that ${\varphi_{5\;}}^3\approx \varphi_{1}$ is the
only degeneracy which occurs in this case, and $\bar V_5$, $ V_1$
are unified into a six-dimensional indecomposable representation
${\mathcal{I}}^{(6)}_{\{5,1\}}$.
Finally the last cases to be considered ($q^4=-1,\;q^3=1$) are not
degenerated, all the vectors (\ref{v35}) distinguish each from
other and all the projection operators $P_5$, $P_3$ and $P_1$ are
well-defined at these points. 
 \bea\label{33}
 V_3\otimes V_3=\left\{
 \ba{ccc}
V_1\oplus {\mathcal{I}}^{(8)}_{\{5,3\}}, & q^2=-1,\\
{\mathcal{I}}^{(6)}_{\{5,1\}} \oplus V_3, & q^3=-1,\\
\;\;\;V_1 \oplus V_3\oplus V_5, & otherwise.
 \ea\right.
 \eea
\paragraph {
$\textbf{V}_3\otimes \textbf{V}_5$.} Fifteen vectors which form
basis of the tensor product
$(\{1,\theta_1,x_1\}\otimes\{1,\theta_2,x_2, x_2\theta_2,x_2^2\})$
are decomposed into the direct sum $V_3\otimes V_5=V_3\oplus
V_5\oplus V_7$ for the general values of $q$,
  \be
c_{3\times5}=[\frac32]^2_q P_3+[\frac52]^2_q P_5+[\frac72]^2_q
P_7.\label{35}
  \ee
Analysis shows that the following degeneracies take place:
$c_7=c_5$, $\{\varphi_7^{a+1}\}\approx\{\varphi_5^a\}$ at $q^3=\pm
1$, $c_7=c_3$, $\{\varphi_7^{i+2}\}\approx\{\varphi_3^{i}\}$ at
$q^5=-1$ and $c_5=c_3$,
$\{\varphi_5^{i+1}\}\approx\{\varphi_3^i\}$ at $q^2=-1$. For these
exceptional cases the conventional spins addition rule (\ref{35})
doesn't work: some homogeneous vectors belonging to different
items in the r.h.s. of (\ref{35}) coincide each to other.
 The fusion rules at any value of $q$
look like:
  \bea
  V_3 \otimes V_5 \quad =\left\{\ba{cc}V_3\oplus
{\mathcal{I}}^{(12)}_{\{7,5\}},
  & \quad q^3=1,\\
{\mathcal{I}}^{(10)}_{\{7,3\}} \oplus V_{5}, & \quad q^{5}=-1,\\
  V_3 \otimes V_5 \otimes V_7, & \quad q^3\neq \pm 1,\;\;q^5\neq -1,\;\;q^2\neq
  -1.
  \ea\right.\label{q35}
  \eea
As we saw in (\ref{33}) the representation $V_5$ absents among the
items in the decomposition of the tensor product $V_3\otimes V_3$
for the values $q^{2}=-1,\quad q^3=-1$. To explain this fact let
us consider what happens with  $V_5$ in the limits $q^{2}=-1$ or
$q^3=-1$. We find that this five-dimensional representation is
non-completely reducible (it has proper subspace which remains
invariant under action of algebra generators). We denote such
representations by $\bar{V}_{d}$, as we did in the previous
examples for the maximal proper sub-representations of
${\mathcal{I}}^{(r)}_{\{k,r-k\}}$-representations. The fusion
rules can be written down for such representations as well.
  \bea\label{i35}
  V_{3} \otimes \bar{V}_5 \quad =\left\{
\ba{cc}
\bar{V}_7\oplus \mathcal{I}^{(8)}_{\{5,3\}}, & q^2=-1,\\
V_3\oplus \mathcal{I}^{(12)}_{\{7,5\}},& q^3=-1.
  \ea\right.
  \eea
  Note that representations $\mathcal{I}^{(12)}_{\{7,5\}}$ for the
cases $q^3=-1$ and $q^3=1$ are different: as we have already
known, the representation $V_5$ is an irrep when $q^3=1$, in
contrast to the first case. And at $q^3=-1$ the representation
$\mathcal{I}^{(12)}_{\{7,5\}}$ has more than two lowest and two
highest weights.

 As it was mentioned (see
also (\cite{KKH})), the matrix representing the Casimir operator
$c$ for the indecomposable representations besides diagonal part
contains also non-diagonalizable triangular blocks, which couple
the eigenvectors with same $k$-values.

To finish this section we would like to consider one more example:
\paragraph {
$\textbf{V}_3\otimes \textbf{V}_3\otimes \textbf{V}_3$.} For
general values of $q$ using the associativity of the tensor
product we can write
  \bea\label{333q}
V_3\otimes V_3\otimes V_3=V_3\otimes(V_1\oplus V_3\oplus V_5)=
V_3\oplus(V_3\otimes V_3)\oplus( V_3\otimes V_5)=V_1\oplus
3(V_3)\oplus 2(V_5)\oplus V_7.
  \eea
Here multipliers 2 and 3 mean multiplicities of the corresponding
items in the decomposition. One can calculate the eigenvectors of
Casimir operator:
\bea\nn & c\cdot\varphi_{1}=[1/2]_q^2\varphi_{1},\qquad
c\cdot\varphi_{7}^\alpha=[7/2]_q^2\varphi_{7}^\alpha,&
\qquad\alpha=1,2,\ldots 7,\\
& c\cdot\varphi_{5}^i=[5/2]_q^2\varphi_{5}^i,\qquad
c\cdot{\varphi^\prime}_{5}^i=[5/2]_q^2{\varphi^\prime}_{5}^i,&
\qquad i=1,2,3,4,5, \label{v333}\\\nn
&\!\!c\cdot\{\varphi_{3}^a,\;{\varphi^\prime}_{3}^a,\;
{\varphi^{\prime\prime}}_{3}^a\}
=[3/2]_q^2\{\varphi_{3}^a,\;{\varphi^\prime}_{3}^a,\;
{\varphi^{\prime\prime}}_{3}^a\},& \qquad a=1,2,3, \eea
where $\varphi_r^a,\; {\varphi^\prime_r}^a,\;
{\varphi^{\prime\prime}_r}^a$ are the eigenvectors corresponding
to the $r$-dimensional representations in the decomposition above.
Taking into account the previous analysis, one expects that the
vectors $\varphi$ become linearly dependent for the same values of
$q$ as for $(V_3\otimes V_3)$ and $(V_3\otimes V_5)$. For the
cases $q^3=1,\;q^5=-1$, when spin-1 irrep $V_5$ exists, from
(\ref{333q}) and (\ref{q35}) it follows
\bea  V_3\otimes V_3\otimes V_3=\left\{\ba{cc} V_1\oplus
3(V_3)\oplus
V_5\oplus \mathcal{I}^{(12)}_{\{7,5\}}, \qquad q^3=1\\
V_1\oplus 2(V_3)\oplus 2(V_5)\oplus \mathcal{I}^{(10)}_{\{7,3\}},
\qquad q^5=-1.\ea\right. \eea
 This rule can be obtained also from the
direct analysis of the eigenvectors (\ref{v333}). The following
degeneracies take place:
  $ q^2\varphi_{7}^{i+1}-\varphi_{5}^{i}+q^{-2}{\varphi^
\prime}_{5}^{i}=0$, when $q^3=1,\; i=1,...,5$ ($q^3=-1,\; i\neq
3$) the spin-$3/2$ and a combination of two spin-$1$
representations are unified into one representation:
$\mathcal{I}^{(12)}_{\{7,5\}}=\widehat{\bar{V}_7\oplus V_5}$, at
$q^3=\pm 1$, and an orthogonal combination is unified with
$\varphi_0$ into
$\mathcal{I}^{(6)}_{\{5,1\}}=\widehat{\bar{V}_5\oplus V_1}$ at
$q^3=-1$.

For $q^5=-1$ the following relation takes place: $
\varphi^{a+2}_{7}\approx(q^{-1}\varphi^{a}_{3}+
{{\varphi^{a}}^\prime}_{3}+q{{\varphi^{a}}^{\prime\prime}}_{3}), $
 which means, that the sum of spin-$3/2$ and a combination of three
spin-$1/2$ irreps transforms into
$\mathcal{I}^{(10)}_{\{7,3\}}=\widehat{\bar{V}_7\oplus V_3}$.

Finally, when $q^2=-1$, we have: $\qquad
\varphi_7^{4}+q\varphi_{1}=0\qquad$ and
\newline
${
}\qquad\qquad\varphi^{a+1}_{5}-(1+q)\varphi^{a}_{3}+{{\varphi^{a}}
^\prime}_{3}+q{{\varphi^{a}}^{\prime\prime}} _{3}=0$,  $\qquad
{\varphi^\prime}^{a+1}_{5}+\varphi^{a}_{3}
-q^{-1}{{\varphi^{a}}^\prime}_{3}
-(1+q){{\varphi^{a}}^{\prime\prime}}_{3}=0$,
\newline
these degeneracies indicate that when $q^2=-1$ the vectors of two
spin-$1$ irreps are unified with two combinations of three
spin-$1/2$ irreps  and produce two $\mathcal{I}^{(8)}_{\{5,3\}}=
\widehat{\bar{V}_5\oplus V_3}$ - representations, and another
indecomposable representation $\mathcal{I}^{(8)}_{\{7,1\}}$ arises
from the unification of spin-$3/2$ and spin-$0$ irreps:
$\widehat{\bar{V}_7\oplus V_1}$.  To say more correct, at the
exceptional values of $q$ the corresponding items in the tensor
product decomposition (\ref{v333}) are replaced by indecomposable
representations.

With respect to the case (\ref{q35})  new features appear only for
the values $q^2=-1,\, q^3=-1$.
  \bea V_3\otimes V_3\!\otimes\! V_3\!=\!
  \left\{\ba{cc}2(\mathcal{I}^{8}_{\{5,3\}})
\! \oplus\! \mathcal{I}^{(8)}_{\{7,1\}}\! \oplus\! V_3,& q^2=-1,
\\\mathcal{I}^{(12)}_{\{7,5\}}
  \!\oplus\! 3(V_3)\!\oplus\! \mathcal{I}^{(6)}_{\{5,1\}}, &  q^3=-1.
\ea\right.\qquad\qquad
  \qquad \label{333}\eea
They provide us with decomposition rules for the tensor products
$V_3\otimes \mathcal{I}^{(8)}_{\{5,3\}}$ and $V_3\otimes
\mathcal{I}^{(6)}_{\{5,1\}}$. Using the associativity property of
the tensor product one can deduce:
  \bea  \mathcal{I}^{(8)}_{\{5,3\}}\!\otimes \!V_{3}=
2(\mathcal{I}^{(8)}_{\{5,3\}}) \oplus
\mathcal{I}^{(8)}_{\{7,1\}},\;\;\;q^2=-1;\qquad\qquad
\mathcal{I}^{(6)}_{\{5,1\}}\!\otimes\! V_3=
\mathcal{I}^{(12)}_{\{7,5\}}\oplus 2(V_3),\;\;\; q^3=-1.
  \label{333i}
  \eea
Here the representation $\mathcal{I}^{(12)}_{\{7,5\}}$ is
decomposable into two six dimensional indecomposable
representations (with the eigenvalues of $k$ respectively
$\{q^{\frac{3}{2}},q,q,q^{\frac{1}{2}},q^{\frac{1}{2}},1\}$ and
$\{1,q^{-\frac{1}{2}},q^{-\frac{1}{2}},q^{-1},q^{-1},q^{-\frac{3}{2}}\}$),
which don't coincide with $\mathcal{I}^{(6)}_{\{5,1\}}$, rather
having the structure of $\mathcal{I}^{(6)}_{\{4,2\}}$ (\ref{23})!

We note, that there is a correspondence between the above
decompositions (\ref{333i}) and the results in (\ref{i35}), which
is the consequence of $\bar{V}_5\subset
\mathcal{I}^{(8)}_{\{5,3\}},\mathcal{I}^{(6)}_{\{5,1\}}$ and
$\bar{V}_7\subset
\mathcal{I}^{(8)}_{\{7,1\}},\mathcal{I}^{(12)}_{\{7,5\}}$. The
only difference is that $\mathcal{I}^{(12)}_{\{7,5\}}\; (q^3=-1)$
is decomposable in (\ref{333i}): this difference arises from the
distinction in the structure of the representation $\bar{V}_5$, as
for the first case we have fixed it by direct choosing the
representation's polynomial space as
 $\{1=f^4\cdot x^2/[2]_q,...f\cdot
x^2/[2]_q\;,x^2\}$, while in the second case $\bar{V}_5$ emerges
from the product $V_3\otimes V_3$ and here
$\Delta(f)^3=0\;(\Delta(e)^3=0)$ at $q^3=-1$.

 One can
also check directly the following fusion for representation
$\mathcal{I}^{(8)}_{\{7,1\}}$, appeared in (\ref{333i}), 
%
  \bea
\mathcal{I}^{(8)}_{\{7,1\}}\otimes
V_3=\mathcal{I}^{(16)}_{\{9,7\}} \oplus
\mathcal{I}^{(8)}_{\{5,3\}},\quad q^2=-1, \label{i8v}
  \eea
where $\mathcal{I}^{(16)}_{\{9,7\}}$ representation arises from
the merging of $V_9$ and $V_7$. This representation consists of
two indecomposable representations in the form of
$\mathcal{I}^{(8)}_{\{5,3\}}$, with the eigenvalues of $k$ being
respectively
$\{q^{2},q^{\frac{3}{2}},q^{\frac{3}{2}},q,q,q^{\frac{1}{2}},q^{\frac{1}{2}},
1\}$ and
$\{1,q^{\frac{-1}{2}},q^{\frac{-1}{2}},q^{-1},q^{-1},q^{\frac{-3}{2}},
q^{\frac{-3}{2}},q^{-2}\}$.

We have seen in the considered examples that non irreducible
representation $\mathcal{I}^{(r)}_{\{k,r-k\}}$, emerging from the
\textit{multiple tensor products} of the irreps, is an
indecomposable representation only in the case, when $V_{r-k}$ is
an irrep for the given $q$.

 Then quartic product of the irreps
$V_3$'s gives decomposition rule for
$\mathcal{I}^{(8)}_{\{5,3\}}\otimes \mathcal{I}^{(8)}_{\{5,3\}}$
when $q^2=-1$:
  \bea\label{ii}
  \otimes^4 V_{3}&=&\otimes^2\! \left(\mathcal{I}^{(8)}_{\{5,3\}}\!\oplus\!
V_{1}\right)=\!\left(\mathcal{I}^{(8)}_{\{5,3\}}\!\otimes\!
\mathcal{I}^{(8)}_{\{5,3\}}\right)\!\oplus\!
2(\mathcal{I}^{(8)}_{\{5,3\}})\! \oplus\! V_{1}\!\\\nn
&=&\!\left(2(\mathcal{I}^{(8)}_{\{5,3\}})\! \oplus\!
\mathcal{I}^{(8)}_{\{7,1\}}\! \oplus \!V_3\right)\!\otimes\!
V_{3}=6(\mathcal{I}^{(8)}_{\{5,3\}})\!\oplus\!
2(\mathcal{I}^{(8)}_{\{7,1\}})\!\oplus\!
(\mathcal{I}^{(16)}_{\{9,7\}})\!\oplus\! V_{1},\\
\!&\Rightarrow &\;\; \mathcal{I}^{(8)}_{\{5,3\}}\!\otimes\!
\mathcal{I}^{(8)}_{\{5,3\}}\!=4(\mathcal{I}^{(8)}_{\{5,3\}})\!\oplus\!
2(\mathcal{I}^{(8)}_{\{7,1\}})\!\oplus\!
(\mathcal{I}^{(16)}_{\{9,7\}})=6(\mathcal{I}^{(8)}_{\{5,3\}})\!\oplus\!
2(\mathcal{I}^{(8)}_{\{7,1\}}).\quad\nn
  \eea
Moreover for the case $q^2=-1$ one can sketch out the fusion of
the tensor product of spin-1/2 representations of number $k$ in
closed form, as follows (see details in the section $5$)
  \bea\label{vvk}
\underbrace{V_3\otimes V_3\cdot\cdot\cdot\otimes V_3}_{k}=\ba{cc}
\bigoplus_{p=1;\;\alpha}^{k/2}\varepsilon_k(p,\alpha)
\mathcal{I}^{(8 p)}_\alpha\oplus V_{1},&\mbox{for even} \quad
k,\\\bigoplus_{p=1;\;\alpha}^{(k-1)/2}
\varepsilon_k(p,\alpha)\mathcal{I}^{(8 p)}_\alpha\oplus
V_3,&\mbox{for odd}\quad k,\ea
  \eea
here $\varepsilon(p,\alpha)$ stands for multiplicity of
$\mathcal{I}^{(8 p)}_\alpha$, where
$\alpha=\{4p+1,4p-1\},\;\{4p+3,4p-3\;\}$.

\section{Tensor product of arbitrary lowest weight irreps
and projectors}

\paragraph{The fusion of two arbitrary irreps $V_n\otimes V_m$.}
The even- and odd-dimensional irreps can be considered on equal
footing.  Suppose we have two finite-dimensional lowest weight
representations $V_n$ and $V_m$ at general values of $q$ and let
for definiteness $n\leq m$. Then eigenvalues of Casimir operator
are built as follows: the lowest weight vectors $\varphi_i$ are
defined as solutions to the equation
 \be\label{lowv}
f\cdot\varphi_i=(f_1+k_1 f_2)\varphi_i=0.
 \ee
The number of these solutions is precisely equal to $n$ because
they are built using $n$ independent vectors of $V_n$. Then each
lowest weight vector gives rise to the invariant subspace of
Casimir operator by successive action of the rising operator
$e
$ (\ref{coprod}) on that vector. The invariant subspace with
largest dimension contains the lowest weight vector $1$ and the
highest one with weight $\frac{n-1}4+\frac{m-1}4$, i.e. has
dimension $n+m-1$. It means that decomposition (\ref{t}) contains
exactly $n$ terms:
 $\qquad
V_n\otimes V_m=V_{m-n+1}\oplus V_{m-n+3}\oplus\ldots\oplus
V_{m+n-1}.\qquad
 $
Consequently the Casimir operator has the following decomposition
over the corresponding projection operators (in accordance with
(\ref{t})):
 \be\label{j1xj2}
c_{n\times m}=c_{m-n+1}P_{m-n+1}+c_{m-n+3}P_{m-n+3}+\ldots+
c_{m+n-1}P_{m+n-1},
 \ee
where $c_r$ is the eigenvalue of Casimir operator on
$r$-dimensional invariant subspace:
 \be\label{ck}
c_{r}=(-1)^{r+1}\frac{q^r+(-1)^r2+q^{-r}}{q^2-2+q^{-2}}=\left\{\ba{cc}
\;[\frac r2]_q^2,\qquad {\rm if}\;\;r\;\;{\rm
is\;\;odd},\\\;[\frac r2+\lambda]_q^2,\quad {\rm if}\;\;r\;\;{\rm
is\;\;even}.\ea\right.
 \ee

 In the expression (\ref{j1xj2}) the projector $P_r$, defined on
the invariant subspace
with given eigenvalue $c_n$ of the Casimir operator $c_{n\times
m}$, can be written as
 \bea\label{p}
P_r=\prod_{p\neq r}\frac{c_{n\times m}-c_p\II}{c_r-c_p};\qquad
\sum_r P_r=\II,\quad P_r P_p=\delta_{r p}P_r.
 \eea

The Casimir eigenvalues $c_r$ (\ref{ck}) coincide each to other
only at the exceptional values of $q$: $c_{r_1}=c_{r_2}$ is
equivalent to the equation
\be ((-q)^{r_1+r_2}-1)((-q)^{r_1-r_2}-1)=0\label{cc}.\ee
 Some
projectors then become ill-defined, having singularities, which is
a sign that when the equation (\ref{cc}) takes place, then the
decomposition (\ref{j1xj2}) is no longer valid, and the spaces
with the ill-defined projectors can be unified into indecomposable
representations. However this condition is necessary but not
sufficient: zeroes in numerator and denominator in (\ref{p}) can
cancel each to other and some projection operators can survive as
it took place in the examples considered above: for the tensor
product of two two-dimensional representations  Casimir operator
at the special points $q=\pm i$ turns to be multiple of the unity
matrix: $c=\frac12{{\II}}$, and both projectors $\quad
 P_1=(c-c_3\II)/(c_1-c_3),\quad P_3=(c-c_1\II)/(c_3-c_1)
$ remain regular. A similar situation occurs in the example
$V_3\otimes V_5$, when $c_3=c_5$ at $q^4=-1$. A careful analysis
shows that at $q^4=-1$ the Casimir operator $c$ satisfies the
relation: $(c - c_7 \II)(c - c_5 \II)|_{q^4=-1}=0, $ and the
projectors $P_3,\;P_5$ survive.

Let us consider $V_n\otimes V_m$ for the cases with ${n=2,\;3}$
separately. \vspace{-0.5cm}
\paragraph{The tensor product of two-dimensional and an arbitrary irrep.}
 The fusion
rule is \be V_2\otimes V_m=V_{m-1}\oplus V_{m+1}, \ee and only two
projection operators exist: $\quad c_{2\times
m}=c_{m-1}P_{m-1}+c_{m+1}P_{m+1}, $
 \be\label{2m}
P_{m-1}=\frac{1}{c_{m+1}-c_{m-1}}(c_{m+1} \II- c_{2\times
m}),\qquad P_{m+1}=\frac{1}{c_{m+1}-c_{m-1}}( -c_{m-1}\II+
 c_{2\times m}). \ee
 From this form
of projection operators one immediately deduces that the only
indecomposable representation which can appear in case of
$c_{m+1}=c_{m-1}$ is ${\mathcal{I}}^{(2m)}_{\{m+1,m-1\}}$, when
$q^{2m}=1$ (\ref{cc}).
\paragraph{The tensor product  of three- and an arbitrary-dimensional irrep.}
 In this case for general  values of $q$ it is valid the
 decomposition
 $$V_3\otimes V_r=V_{r-2}\oplus V_r\oplus
V_{r+2}$$ and one has: $ c_{3\times
r}=c_{r+2}P_{r+2}+c_{r}P_{r}+c_{r-2}P_{r-2}, $ where
the structure of the denominators in the expressions of $P_r$ (see
(\ref{p})) suggests that the projectors can be singular when $c_r=
c_{r-2}$, $c_{r+2}=c_{r-2}$ and/or $c_{r+2}= c_r$. As we shall see
only two kind of indecomposable representations can appear in this
fusion: ${\mathcal{I }}^{2r}_{\{r+2,r-2\}}$
($c_{r+2}=c_{r-2},\;q^{2r}=1,q^4=1$) and
${\mathcal{I}}^{2r+2}_{\{r+2,r\}}$($c_{r+2}=c_{r},\;q^{2r+2}=1$).

\section{General results: 
Fusion rules }

The aim of this section is to clarify the peculiarities of the
finite dimensional representations and their fusions which occur
at exceptional values of $q$, $q^N=\pm 1$, for general $N\in
\mathbb{N}$. 
The considered examples show that the number of irreducible
representations is restricted, when $q$ is given by a root of
unity, and the new type of representations - indecomposable
representations, appears in the fusions. Is it possible to extend
observed regularities to general $N\in\mathbb{N}$, finding all
finite dimensional non reducible representations with their fusion
rules, and the relations between $N$ and the dimensions of the
permissible representations?

As we have already seen above, when $q^N=\pm 1$, the irrep $V_r$,
since $r>r_{max}$,  becomes non-irreducible representation $\bar
V_r$, which contains one or more proper subspaces. Such
representations do not appear in the fusions of the irreps, but
indecomposable representations appearing in the tensor products'
decompositions contain such representations as sub-representations
($\mathcal{I}=\widehat{\bar{V}\oplus V}$). This observation allows
us to trace the connection between the number $N$ and the
dimensions of the permissible irreps (i.e. $r_{max}$) and the
indecomposable representations.

All the  representations can be constructed uniformly, in a
general form. As usual, one can choose as  basis vectors of a
representation the eigenvectors $|h_n\rangle$ of the operator $k$.
 \be
k |h_n\rangle=k_n|h_n\rangle, \quad k_n=q^{h_n}\in\mathbb{C}.
 \ee
Then from the algebra relations ($\!$\ref{alg}$\!$) one obtains
constraints on the actions of the operators $e$ and $f\!$:
 \bea \nn
k(e^m|h_n\rangle)=q^m k_n(e^m|h_n\rangle),\qquad
k(f^m|h_n\rangle)=q^{-m}k_n(f^m|h_n\rangle).
 \eea
If $e^m|h_n\rangle\neq 0$ and $f^m|h_n\rangle\neq 0$, then
$e^m|h_n\rangle\approx|h_n+m\rangle$ and $f^m|h_n\rangle\approx
|h_n-m\rangle$ are also the eigenvectors of $k$-operator, with the
eigenvalues of $k$ being the powers of $q$, $ q^{h_n \pm m}$. But
for $q^N=1$ the spectrum of the eigenvalues of the operator $k$
gets degenerated: the states $|h_n\rangle,\quad|h_n\pm
N\rangle,\quad|h_n\pm2N \rangle,...$ have the same eigenvalue of
$k$. It means that in this case one has:
\be\label{N} f|h_n\rangle\!=\!\alpha_1
|h_n-1\rangle+\!\alpha_2|\!h_n-\!1\pm\!N\!\rangle+\cdots,\;\;
e|h_n\rangle\!=\!\alpha'_1 |h_n+\!1\!\rangle+\!\alpha'_2|h
_n+1\pm\!N\!\rangle+\cdots. \ee
The parameters $\alpha_i,\; \alpha'_i$ define the representation,
and the anti-commutation relation between $e$ and $f$ imposes
constraints on them. Different values of these parameters
correspond to reducible or non-reducible representations (cyclic,
semi-cyclic, nilpotent or lowest/highest weight ones
\cite{JimGRAS}).
 In particular, at general values of $q$ the
finite dimensional irreps can be found suggesting the existence of
the lowest weight vector $|h_0\rangle,\quad f|h_0\rangle=0$. Then
the following relation takes place
 \be\label{h0}
fe^r|h_0\rangle=\left([h_0+r-1]_q-[h_0+r-2]_q+\cdots +
 (-1)^{r-1}[h_0]_q\right)e^{r-1}|h_0\rangle.
 \ee
If the r.h.s vanishes, then the representation $\{|h_0\rangle,
\;|h_1\rangle=f|h_0\rangle,\;\;...\;|h_r\rangle=e^{r-1}|h_0\rangle\}$
is an $r$-dimensional irreducible lowest weight (by construction)
representation, and the possible values of $h_0$ can be obtained
from the analysis of the zeros of (\ref{h0}), which gives
$q^{2h_0}=(-1)^{r-1}q^{1-r}$. For odd values of $r$, the
eigenvalues $h_p$ take integer values ($h_p \in \{
(1-r)/2,(3-r)/2,...,(r-1)/2 \}$, $q$-analog of the conventional
spin irreps with spin ${(r-1)/4}$), while for even dimensional
irreps the values $h_0$ contain the nontrivial term
$(\imath\pi/(2\log{q}))$ \cite{Ki}, see
 (\ref{lambda}).
For the exceptional values of $q$, as it was already mentioned in
the second section, the lowest/highest weight representations
emerged from the fusions of the fundamental spin-half irreps are
distinguished by the values equal to $0$ of the operators
$e^{\mathcal{N}},\;f^{\mathcal{N}}$, where $\mathcal{N}=\left\{
{}^{N,\;even\; N}_{2N,\;odd\; N}\right. $ if $q^N=1$ or
$\mathcal{N}=\left\{{ }^{N,\; odd\; N}_{2N,\; even \; N}\right.$
if $q^N=-1$. 
%
%
%
%


\subsection{Odd dimensional conventional representations and indecomposable
representations}

The odd dimensional representations for general values of $q$ form
a closed fusion (\ref{t})
 \be
V_{4j_1+1} \otimes V_{4j_2+1}=\bigoplus_{j=|j_1-j_2|}^
{j_1+j_2}V_{4j+1},\quad\qquad\quad\Delta j=\frac{1}{2},
\label{v*v}
 \ee
$j$ is integer or half-integer. In this part we are considering
only odd-dimensional representations and their fusions at roots of
unity, but the whole analysis can be carried out with the
inclusion of the even-dimensional ones as well. It is presented in
the next section. 

\paragraph{Representation $V$.} For the general values of $q$, the action of the generators
$e$, $f$ and $k$ on the vectors $V_{4j+1}=\{v_j(h)\}$ of the
\textit{odd dimensional spin-$j$ representation} can be written as
 \bea\label{odd} \hspace{-3cm}\left\{
 \ba{lll}f\cdot v_j(h)=\gamma_h^{j}(q) v_j(h-1),&
 -2j<
h\leq 2j,\quad &f \cdot v_j(-2j)=0,\\
k\cdot v_j(h)=q^{h} v_j(h),&-2j\leq h\leq 2j,&\\
 e\cdot
v_j(h)=\beta_h^{j}(q)v_j(h+1),&  -2j\leq h<2j,\quad &e \cdot
v_j(2j)=0,\ea
 \right.
 \\\nn
\gamma_h^{j}(q)\beta_{h-1}^{j}(q)=\alpha^{j}_h(q).\hspace{3cm}\eea
The algebra relations imply the following expressions for the
coefficients $\alpha^{j}_h(q)$
 %
%
\bea
\label{alpha}\alpha^{j}_{h}(q)=\sum_{i=h}^{2j}(-1)^{i-h}[i]_q&=&\frac
{(-1)^{2j+h}[2j+1/2]_q+[h-1/2]_q}{\sqrt{q}+1/\sqrt{q}}, \quad
-2j<h\leq 2j,\\
\alpha^{j}_{h}(q)&=&-\alpha^{j}_{-h+1}(q),\quad -2j<h\leq 0.
 \eea
Usually the coefficients $\beta_h^{j}(q), \; \gamma_h^{j}(q)$ are
chosen imposing some normalization conditions on the basis vectors
(e.g. defining a norm \cite{S}, such that $\langle
v_j(h)|v_j(h)\rangle=1$, if the lowest weight vector $v_j(-2j)$
has parity $0$, and $\langle v_j(h)|v_j(h)\rangle=1(-1)$ for
$2j-h$ even integer (odd integer), if the lowest weight vector has
parity $1$; and setting $f$ to be the adjoint of $e$, when
 the adjoint of an operator $g$ is defined as $\langle g^*\cdot v|u
\rangle=(-1)^{p(v)p(u)}\langle v|g\cdot u \rangle$).
 But it is failed when $q$ is given by a root of unity.

 Here
  $\beta_h^{j}(q), \; \gamma_h^{j}(q)$ define the sub-structure of
the representation. However the possible choice does not affect
the conclusions given below for the fusion rules. One can
formulate the following \vspace{-5mm}
\paragraph{{Statement I:}}
\textit{the representation $V_{4j+1}$ contains invariant
sub-representations, if at least one of the functions
$\alpha^{j}_{h}(q),\;  -2j< h\leq 2j\;$, describing  $V_{4j+1}$ is
equal to zero}.

If $\alpha_h^j(q)=0$, then we call the representation $V_{4j+1}$
non-exactly-reducible and denote it by $\bar{V}_{4j+1}$. This
representation is not irrep and contains more than one highest and
more than one lowest weight vectors (which can be
$v_{j}(\pm(h-1))$).

 If the functions $\beta_h^{j}(q),\;\gamma_h(q)$ in the definition
(\ref{odd}) are chosen as
 \be
\beta_{h-1}^{j}(q)=1,\;\;\gamma_h^{j}(q)=\alpha^{j}_{h}(q),\;-2j <
h < 1 ,
\;\;\beta_{h-1}^{j}(q)=\alpha^{j}_{h}(q),\;\;\gamma_h^{j}(q)=1,\;\;
1\leq h \leq 2j, \label{ef}\ee
then $\alpha^{j}_{2j'+1}(q)=0$ ($j'>0$) indicates the appearance
of the invariant sub-representation $\{v_j(h)\}$, $\; -2j'\leq
h\leq 2j'$ inside of $\bar{V}_{4j+1}$. In the figure
(Fig.\ref{fr}b) we described  representation
$\bar{V}_{4j_2+1}\supset V_{4j_1+1}$ diagrammatically, denoting
states $v_j(h)$ by dots  (the corresponding values of $h$ are
noted at the left column). On the diagram the arrows $\uparrow$
and $\downarrow$ correspond to the action of the raising and
lowering operators. In (Fig.\ref{fr}) all the dots, that are not
shown on the diagrams, are connected with their nearest neighbors
with both arrows ($\uparrow$ and $\downarrow$).  In this case
there are two highest weight vectors, $v_{j_2}(2j_2), \;
v_{j_2}(2j_1)$, and two lowest weight vectors, $v_{j_2}(-2j_2), \;
v_{j_2}(-2j_1)$. For the cases, when $\bar{V}_{r}$ has more than
two highest and two lowest weight vectors, we should depict the
diagram for $\bar{V}_{r}$ in a similar way, omitting the
$\uparrow$-arrows, connected the dots describing the highest
weight vectors with their upper nearest neighbors, and
$\downarrow$-arrows, connected the dots of the lowest weight
vectors with their lower nearest neighbors.

Note, that if one chooses $\beta_h^{j}(q),\;\gamma_h^{j}(q)$ to be
proportional to $\sqrt{\alpha_{h+1}^{j}(q)},\;\;
\sqrt{\alpha_h^{j}(q)}$ (imposing $e=f^{\tau}$), then
representation $\bar{V}_{4j+1}$ will be completely reducible when
some $\alpha_{h}^{j}(q)=0, \quad -2j< h\leq 2j$ (e.g. in the
example described in Fig.\ref{fr}b, $\bar{V}_{4j_2+1}$ would be
split into one (${4j_1+1}$)- and two $2(j_2-j_1)$-dimensional
representations).

It follows from (\ref{alpha}) that $\alpha^{j}_{h}(q)=0$ is
equivalent to the equation
 \be \label{cc1}
(1-(-1)^{2j+h}q^{2j+h})(1+(-1)^{2j-h}q^{2j+1-h})=0.
 \ee
Taking into account that
$\alpha^{j}_{h}(q)=-\alpha^{j}_{-h+1}(q)$, 
 we can consider only the solutions to the equation
$ q^{2j+h}=(-1)^{2j+h}$,
which, for the whole range of the eigenvalues $h$, $-2j < h \leq
2j$, can take place if $ q^n=1$, with $n$ is an even integer, or
if $q^n=-1$, with $n$ is odd, and at the same time $2j+h=n p$, $p$
is positive integer, with the range in the interval $1\!<n p\leq
4j$.

We can summarize as follows: $\alpha^{j}_{h}(q)=0$, if
\bea\label{qnn}
 \left\{\ba{cc}q^{N}=\pm 1,\quad & N \quad \mbox{is even},\\
q^{N}=1,\quad & N\quad  \mbox{is odd},\ea\right.
\quad\quad\mbox{with}&
&h=2N p-2j, \quad\quad\ba{c} \\ \ea\\
\hspace{-1cm}q^{2N-1}=-1,\hspace{1cm}\qquad \qquad \qquad
\mbox{with}& &h=(2N-1)p-2j.\label{qn} \eea
Here $1<N\; p\leq 2j$, and the case, when $q^N=1$ and  $N$ is
even, could be omitted, as it is equivalent to the case
$q^{N/2}=-1$.

So, if $q$ satisfies one of the relations (\ref{qnn}, \ref{qn}),
then for the corresponding $\{j,\;h\}$-s one has
$\alpha^{j}_{h}(q)=0$ and $V_{4j+1}$ is no longer irreducible and
should be denoted as $\bar{V}_{4j+1}$.

On the other hand, from (\ref{qnn}, \ref{qn}) it follows, that for
a given $N$, the \textit{permissible} irreps are the
representations $V_{4j+1}$ with spin $j$, which satisfies the
inequality
%
%
\bea \ba{c}{\vspace{1cm}}\\j\leq j_{max},\ea \quad
&&j_{max}=\frac{N-1}{2} \qquad
\mbox{for}\qquad\left\{\ba{c}q^N=1,\;\;N\; \mbox{is odd}\\q^N=-1,
\;\;N\; \mbox{is even}\ea \right. \label{jmax1}\\
 &&j_{max}=\frac{N-1}{4}
\qquad \mbox{for}\qquad \quad q^N=-1, \;\; N\; \mbox{is odd}.
\label{jmax2}
 \eea
As we have already seen, $\bar{V}_{4j+1}$-representations,
$j>j_{max}$, do not emerge in the fusions of the irreps, instead
new indecomposable representations appear. Let us summarize
observed regularities
 as \vspace{-5mm}
\paragraph{{Statement II:}} {\textit{
The following three criteria describe the appearance of an
indecomposable representation: 
when in the r.h.s. of the decomposition (\ref{v*v}) any two
representations
$$
V_{4j+1}=\{v_{j}({-2j}),...,v_{j} ({2j}) \}, \quad
V_{4s+1}=\{v_{s}({-2s}),...,v_{s} ({2s}) \},\qquad s<j,
$$
\begin{enumerate}
\item  have the same eigenvalues of the Casimir operator,
$c_{4s+1}=c_{4j+1}$ - necessary criterion,
\item the following eigenvectors of the Casimir operator are
linearly dependent: $v_{s}(h)\approx
v_{j}(h),\;\;\;h\in(-2s,...,2s)$ - necessary and sufficient
criterion,
\item  $\bar{V}_{4j+1}\supset V_{4s+1}$ (i.e. $V_{4j+1}$ turns
into $\bar{V}_{4j+1}$ one), and $\bar{V}_{4j+1}$ has no larger
proper sub-space than $V_{4s+1}$ - necessary and sufficient
criterion,
\end{enumerate}
}}%
{\textit{
it means that the sum $V_{4s+1}\oplus V_{ 4j+1}$ degenerates and
after completion by new vectors $v'(h)$, with the eigenvalues of
generator $k$ being $q^h,\;h\in(-2s,...,2s)$, turns into the
indecomposable representation $\mathcal{I}^{(4
(s+j)+2)}_{\{4j+1,4s+1\}}=\widehat{\bar{V}_{4j+1}\oplus
V_{4s+1}}$.}}


$\it{1})$  It is easy to see, that when the first point does not
take place, then all the spins $j$ in (\ref{v*v}) are
"permissible" (\ref{jmax1}, \ref{jmax2}) and hence the
decomposition (\ref{v*v}) remains unchanged. But it is possible a
situation, when all the spins are "permissible" but a casual
degeneration of the eigenvalues of the Casimir operator takes
place. So, the first point is the simplest necessary, but not
sufficient criterion for the distortion of the usual decomposition
rule.


{{$\it{2})$ The realization of the second point means that the
mentioned vectors belonging to different representations coincide
each to other, so 
the usual decomposition rule (\ref{v*v}) is spoiled. Moreover,
such coincidence of the Casimir eigenvectors from different
multiplets immediately implies coincidence of the corresponding
eigenvalues, i.e the first point follows from the second one. As
$V_{4s+1}$ is an irrep, it means that the vectors $\{v_j(-2s),...,
v_j(2s)\}$ constitute a proper sub-space of the representation
$V_{4j+1}$ (so, the third point realizes as well), and
consequently $j>j_{max}$.

${\it 3} )$ 
Third point implies that the equation $\alpha_{2s+1}^{j}(q)=0$
takes place, and hence $(-q)^{2j+2s+1}=1$ (Statement I and
(\ref{cc1})), i.e. $j>j_{max}$.
  Note, that the solutions to the equations
(\ref{cc1}) are also the solutions to (\ref{cc}), when
$r_1=4j+1,\; r_2=2h-1$, so if for some exceptional $q$ the third
point of the Statement II takes place, the first point is also
true. The relation $\alpha_{h}^{j}(q)=\alpha_{h}^{s}(q)$ is
fulfilled as well, when $(-q)^{2j+2s+1}=1$. Hence the
($4s+1$)-dimensional sub-representation of $V_{4j+1}$ and the
representation $V_{4s+1}$ have the same characteristics. It is
easy to verify, that any linear superposition of the vectors
$v_{j}(h)$ and $v_{s}(h)$ with the weights $h\in(-2s,...,2s)$
belongs (up to numerical coefficients) either to the
representation $V_{4j+1}$ or to $V_{4s+1}$, 
which indicates that the mentioned vectors are linearly dependent
(i.e. the second point follows from the third one too).

And vice versa, any destruction of the Clebsh-Gordan decomposition
at roots of unity means, that there must be a spin $j$ in
(\ref{v*v}) which is larger than $j_{max}$.
 Then for such representation
$V_{4j+1}$ the relation $\alpha_{h'+1}^{j}(q)=0$ takes place
(Statement I) for some $h'$, and consequently $(-q)^{2j+h'+1}=1$
according to (\ref{qnn},
 \ref{qn}).}}
  And as now $\Delta(e^{2j+h'+1})=0,\;
\Delta(f^{2j+h'+1})=0$ (recalling definition of $\mathcal{N}$ and
the formulas (\ref{fN}), (\ref{qnn}, \ref{qn})), so
$\beta^{j}_{-h'}=\gamma^{j}_{h'}=0$, which means that
$\bar{V}_{4j+1}$ has $2h'+1$-dimensional proper
sub-representation. This brings to the situation described in the
third point of the Statement II, with $h'\equiv 2s$, i.e. 
any distortion of the standard fusion rules leads to fulfillment
of the third point, and consequently to the first and
the second points as well.  


 Let us now see that the coincidence of the eigenvectors (${\it 2.}$) leads to the appearance
 of the indecomposable representation. Indeed, as we know, in the decomposition (\ref{v*v}) at
general $q$ the eigenvectors of Casimir operator $v_j(h)$ (in
r.h.s of the equation) make a basis in the space of the tensor
product (l.h.s of the equation), formed by $v_{j_1}(h_1)\otimes
v_{j_2}(h_2)$. The second point shows 
that the number of
 non-zero eigenvectors is reduced (some eigenvectors are
identical to others). 
Hence it is necessary to supplement them with new vectors $v'(h)$
to span the whole space of the decomposition. 
%
  In order to find the vectors $v'(h)$ we can borrow the concept of
the vectors with null norm from the article \cite{PS} (see also
references therein), where it was observed, that when $v_{j}(\pm
2s)$ ($j>s$) are highest and lowest weight vectors, then all the
states $ v_{j}(h),\;\;h\in(-2s,...,2s)$ have null norms. As it was
mentioned already, a norm can be defined in the graded space by
means of a scalar product $\langle v_1|v_2 \rangle$, defining $f$
as the adjoint of $e$. And we can see that $\langle
v_{j}(h)|v_{j}(h) \rangle\approx \langle f^{2s+1-h} \cdot
v_{j}(2s+1)|v_{j}(h) \rangle\approx\langle v_{j}(2s+1)|e\cdot
v_{j}(2s) \rangle=0$,
 $h\in(-2s,...,2s)$.
In the decomposition at general $q$ the vectors $
v_{j}(h),\;\;h\in(-2s,...,2s)$ are orthogonal to the vectors,
belonging to the representation $V_{4s+1}$. But now the pointed
$v_{j}(h)$ are self-orthogonal and  are linearly dependent with
the vectors of $V_{4s+1}$ with same values of $h$. As the
orthogonal space of the non-zero vector $v_{j}(2s)$ contains
itself already, there must exist a state $v'(s)$, with $h=2s$,
which is not orthogonal to $v_{j}(2s)$. It follows from $\langle
v_{j}(2s+1)|e\cdot v'(2s) \rangle \approx \langle f\cdot
v_{j}(2s+1)|v'(2s)\rangle\approx\langle v_{j}(2s)|v'(2s)
\rangle\neq 0$ that $e\cdot v'(2s)= a\;v_{j}(2s+1)$ ($a$ is a
numerical non-zero coefficient). Solving the last equation, and
then acting by $f^{2s-h}$
 on $v'(2s)$, we can find out the remaining states $v'(h),\;h
\in(-2s,...,2s)$, which together with $v_{j}(h),\;h\in(
-2j,...,2j)$, constitute the representation $\mathcal{I}^{(4(
s+j)+2)}_{\{2j+1,2s+1\}}$ (see (\ref{ind})).

{{ So, under the conditions 
 of the Statement II
  a  modification of the
decomposition rule (\ref{v*v}) at roots of unity  can take place
characterized with appearance of $\mathcal{I}$, 
which means that decomposition contains a
representation 
 with 
$j>j_{max}$, and
this 
in it's turn means fulfilment of the mentioned interrelated points. 
We see that the second and third points (which are equivalent each
to other) provide necessary and sufficient criteria for such
distortion, while the first one is only necessary.
}}$\Box$


The points of the Statement II for $V_{r_{max}}\otimes V_3$ are
considered in
details in the Appendix. 

The 
Statements I, II
help us to determine all the possible
$\mathcal{I}$-representations at $q^N=\pm 1$ and to formulate the
modified fusion rules (see the next subsection).


\paragraph{Representation $\mathcal{I}$.} Taking into account its origin
from the fusion we can define \textit{indecomposable
representation} $\mathcal{I}^{(4(j_1+j_2 )+2)}_{\{4j_{2}+1,
4j_{1}+1\}}$ as a linear space $\{\upsilon(h),-2{j_2}\leq h\leq
2{j_2},\;\; \upsilon'(h'),-2{j_1}\leq h'\leq 2{j_1}\},$ with the
following action of the algebra generators:
\bea \hspace{-1cm} \left\{\ba{ll}e \cdot
\upsilon(h)=\beta_{h}^{j_2}(q)\upsilon(h+1),&
 e \cdot \upsilon(2j_1)=0,\;e \cdot
\upsilon(2j_2)=0,\\
 k\cdot \upsilon(h)=q^{h} \upsilon(h),& \qquad
 \\
f\cdot\upsilon(h)=\gamma_{h}^ {j_2}(q) \upsilon(h-1),&
f\cdot\upsilon(-2j_1)=0,\; f\cdot\upsilon(-2j_2)=0,\\
 e\cdot \upsilon'(h')=\bar{\beta}_{h'}^{j_1}(q)\upsilon'(h'+1)
+\tilde{\beta}_{h'}^{j_1}(q)\upsilon(h'+1),&\;\bar{\beta}_{2j_1}
^{j_1}(q)=0,\\
k\cdot\upsilon'(h)=q^{h'} \upsilon'(h'),
 &\qquad
\\
f\cdot \upsilon'(h')=\bar{\gamma}_{h'}^{j_1}(q)\upsilon'
(h'-1)+\tilde {\gamma}_{h'}^{j_1}(q)\upsilon(h'-1),&\;\bar
{\gamma}_{-2j_1}^{j_1}(q)=0,\ea\right.\label{ind}
 \eea
 with
 $\alpha_{2j_1+1}^{j_2}(q)=0$, and
at the same time $2j_1$ is the biggest $h$, for which
$\alpha_{h+1}^{j_2}(q)=0$. Hence the spins $j_1$ and $j_2$ are
related by the equations (\ref{qnn}) and (\ref{qn}), which impose
constraints on $j_1$ and $j_2$, in particular $2(j_2-j_1)\geq 1$.

 New functions
$\bar{\beta}_{h}^{j_1}(q),\;\tilde{\beta}_{h}^{j_1}
(q),\;\bar{\gamma}_h^{j_1}(q),\tilde{\gamma}_h^{j_1}(q)$ are
constrained by the algebra relations, which give
\bea
\bar{\beta}_{h-1}^{j_1}\bar{\gamma}_h^{j_1}={\alpha}_h^{j_1},\quad
\bar{\beta}_h^{j_1}\tilde{\gamma}_{h+1}^{j_1}+
\gamma_{h+1}^{j_2}\tilde{{\beta}}_h^{j_1}+
\bar{\gamma}_h^{j_1}\tilde{{\beta}}_{h-1}^{j_1}+
\beta_{h-1}^{j_2}\tilde{\gamma}_h^{j_1}=0.\eea
So, this representation has the structure described in (\ref{N}).
For general values of $q$ the representation (\ref{ind}) would be,
of course, completely reducible to the direct sum of the irreps
$V_{4j_1+1}$ and $V_{4j_2+1}$.

In the figure (Fig.\ref{fr}a) we presented a general
representation $\mathcal{I}^{(4(j_1+j_2)+1)}_{\{4j_1+1,4j_2+1\}}$
diagrammatically, denoting by dots the states
$\upsilon(h),\;\upsilon'(h)$ (the corresponding values of $h$ are
noted at the left column). The arrows $\uparrow,\;\nwarrow$ show
the action of the raising operator, while the arrows
$\downarrow,\;\swarrow$ correspond to the action of the lowering
operator. In the examples considered in the third section the only
transition ($\nwarrow$) we met was corresponding to the action
$e\cdot \upsilon'(2j_1)=\tilde
{\beta}_{2j_1}^{j_1}(q)\upsilon(2j_1+1)$. It is conditioned by the
fact, that $\upsilon'(h)$-states, with
$h=\!-2j_1\!+\!1,-2j_1\!+\!2,...,$ were obtained by the action on
the state
 $\upsilon'(-2j_1)$ of the operators $e^p,\;p=1,...,4j_1$. Redefining states $\upsilon'(h)$
as $a \upsilon'(h)+b \upsilon(h)$ ($a,\;b\in \mathbb{Z}$), we
should come to the more general case (\ref{ind}).

For a given $N$, $q^N=\pm 1$, the possible dimensions of the
representations $\mathcal{I}^{(4(j_1+j_2)+2)}_{\{4j_2+1,4j_1+1\}}$
can be obtained from (\ref{qnn}, \ref{qn}) with $j=j_2, \; h=2j_1+1$%
: as the dimension of the representation (\ref{ind}) is
$4(j_1+j_2)+2$, so for the integers $N,\; p$ ($1<N p\leq 2j_2$),
we obtain
\be
\mbox{dim}[\mathcal{I}^{(4(j_1+j_2)+2)}_{\{4j_2+1,4j_1+1\}}]=\!
4(j_1+j_2)+2\!=\!\left\{\ba{ll}4 N p,& q^N=-1, \;\;N\;\; \mbox{is
even},\; \&\;\; q^N=1,\;\; N \;\;\mbox{is odd},\\
2N p, &\quad\quad\quad q^N=-1,\;\; N\;\;\mbox{is odd
integer}.\ea\right.\label{in} \ee
%

 For an illustration of the structure of $\mathcal{I}$ we can consider for example the
indecomposable representation $\mathcal{I}^{(8)}_{\{5,3\}}$ at
$q^4=1$ ($q^2=-1$) 
in a basis $\{\upsilon_{h},\; \upsilon'_{h'}\}$ as follow
\begin{small}
\bea\nn
e\cdot\{\upsilon_{2},\upsilon_{1},\upsilon_{0},\upsilon_{-1},
\upsilon_{-2},\upsilon'_{1},\upsilon'_{0},\upsilon'_{-1}\}&=&\{0,0,
-i\upsilon_{1},-i\upsilon_{0},-i\upsilon_{-1},-\upsilon_{2},
\upsilon'_{1},\upsilon'_{0}\},\\\nn f\cdot\{\upsilon_{2},\upsilon
_{1},\upsilon_{0},\upsilon_{-1},\upsilon_{-2},\upsilon'_{1},
\upsilon'_{0},\upsilon'_{-1}\}&=&\{-i\upsilon_{1},i\upsilon_{0},-i
\upsilon_{-1},0,0,\upsilon'_{0}\!+\!\upsilon_{0},-\upsilon'_{-1}
\!-\!i\upsilon_{-1},-\upsilon_{-2}\},\\\nn k\cdot\{\upsilon_{2},
\upsilon_{1},\upsilon_{0},\upsilon_{-1},\upsilon_{-2},\upsilon'_{1}
,\upsilon'_{0},\upsilon'_{-1}\}&=&\{-\upsilon_{2},i\upsilon_{1},
\upsilon_{0},\!-i\upsilon_{-1},\!-\upsilon_{-2},i\upsilon'_{1},
\upsilon'_{0},\!-i\upsilon'_{-1}\}. \!\!\!\!\!\!\! \eea
\end{small}
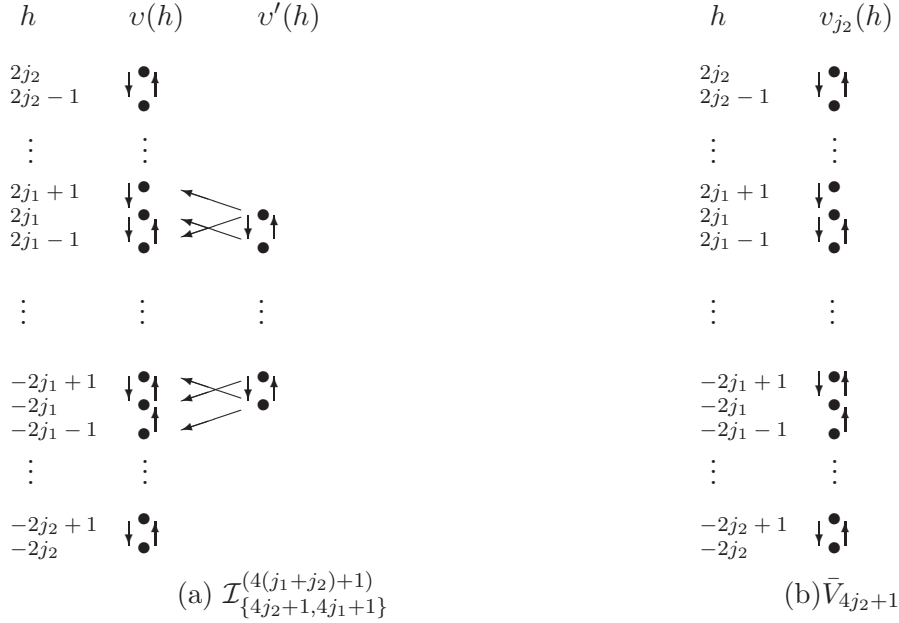
\begin{figure}[t]
\hspace{1 cm} \unitlength=9pt
\begin{picture}(100,23)(12,0)

\put(20,24.3){$\upsilon(h)$} \put(25,24.3){
$\upsilon'(h)$}
\multiput(20,22)(0,-6){2}{$\;\bullet$}

\multiput(20,20.6)(0,-3.4){2}{$\;\bullet$}

\multiput(20,8)(0,-6){2}{$\;\bullet$}

\multiput(20,14.6)(0,-5.4){2}{$\;\bullet$}

\multiput(20,6.8)(0,-3.6){2}{$\;\bullet$}
\multiput(25,16)(0,-8){2}{$\;\bullet$}
\multiput(25,14.6)(0,-5.4){2}{$\;\bullet$}
\multiput(20.5,18.5)(0,-13.5){2}{$\vdots$}
\multiput(15.4,11.72)(5,0){3}{$\vdots$}

\put(15,24.3){ $h$}\put(15,22){\scriptsize $2j_2$}
\put(15,16){\scriptsize$2j_1$}
\put(15,8){\scriptsize $-2j_1$} \put(15,2){\scriptsize $-2j_2$}
\put(15,7){\scriptsize $-2j_1-1$} \put(15,3){\scriptsize
$-2j_2+1$} \put(15,9){\scriptsize $-2j_1+1$}
\put(15,15){\scriptsize $2j_1-1$}\put(15,17){\scriptsize $2j_1+1$}
\put(15,21){\scriptsize $2j_2-1$}
\multiput(15.7,18.5)(0,-13.5){2}{$\vdots$}

\put(24.7,8.1){\vector(-3,-1){2.5}}\put(24.7,16.5){\vector(-3,1){2.5}}
\put(24.7,8.6){\vector(-3,1){2.5}}\put(24.7,16.2){\vector(-3,-1){2.5}}
\put(24.7,9.3){\vector(-3,-1){2.5}}\put(24.7,15.3){\vector(-3,1){2.5}}
\put(26.1,8.5){\vector(0,1){1}}\put(25,9.5){\vector(0,-1){1}}
\put(26.1,15.3){\vector(0,1){1}}\put(25,16.3){\vector(0,-1){1}}
\put(21.1,8.5){\vector(0,1){1}}\put(20,9.5){\vector(0,-1){1}}
\put(21.1,15.1){\vector(0,1){1}}\put(20,16.2){\vector(0,-1){1}}
\put(21.1,7.2){\vector(0,1){1}}\put(20,17.6){\vector(0,-1){1}}

\put(21.1,21.3){\vector(0,1){1}}\put(20,22.3){\vector(0,-1){1}}

\put(21.1,2.4){\vector(0,1){1}}\put(20,3.4){\vector(0,-1){1}}

\put(22,0){(a) $\mathcal{I}^{(4(j_1+j_2)+1)}_{\{4j_2+1,4j_1+1\}}$}


\put(49,24.3){$v_{j_2}(h)$} \put(44,24.3){ $h$}
\put(44,22){\scriptsize $2j_2$}

\multiput(49,22)(0,-6){2}{$\;\bullet$}

\multiput(49,20.6)(0,-3.4){2}{$\;\bullet$}

\multiput(49,8)(0,-6){2}{$\;\bullet$}

\multiput(49,14.6)(0,-5.4){2}{$\;\bullet$}

\multiput(49,6.8)(0,-3.6){2}{$\;\bullet$}
\multiput(49.5,18.5)(0,-13.5){2}{$\vdots$}
\multiput(44.4,11.72)(5,0){2}{$\vdots$}
\multiput(44.4,18.5)(0,-13.5){2}{$\vdots$}

\put(44,8){\scriptsize $-2j_1$} \put(44,2){\scriptsize $-2j_2$}
\put(44,7){\scriptsize $-2j_1-1$} \put(44,3){\scriptsize
$-2j_2+1$} \put(44,9){\scriptsize $-2j_1+1$}
\put(44,15){\scriptsize $2j_1-1$}\put(44,16){\scriptsize
$2j_1$}\put(44,17){\scriptsize $2j_1+1$} \put(44,21){\scriptsize
$2j_2-1$}

\put(50.1,8.7){\vector(0,1){1}}\put(49,9.7){\vector(0,-1){1}}
\put(50.1,15.1){\vector(0,1){1}}\put(49,16.2){\vector(0,-1){1}}
\put(50.1,7.2){\vector(0,1){1}}\put(49,17.6){\vector(0,-1){1}}

\put(50.1,21.3){\vector(0,1){1}}\put(49,22.3){\vector(0,-1){1}}

\put(50.1,2.4){\vector(0,1){1}}\put(49,3.4){\vector(0,-1){1}}

\put(47.5,0){(b)$\bar{V}_{4j_2+1}$}
\end{picture}
\caption{Representations (a)
$\mathcal{I}^{(4(j_1+j_2)+1)}_{\{4j_1+1,4j_2+1\}}$ and
(b)$\bar{V}_{4j_2+1}\supset V_{4j_1+1}$}\label{fr}
\end{figure}
\vspace{-1cm} 
\paragraph{{sdim}$_q$.} The notion of $q$-superdimension
(for the non-graded algebras - $q$-dimension) \cite{PS, S, KKH} of
the representation $V$, $\;\textmd{sdim}_q (V)=\verb"str" \; k$,
where $\verb"str"$ denotes super-trace defined in the graded space
of the representation, will be useful here. For the
representations $V_{4j+1}$ (or $\bar{V}_{4j+1}$)
\bea\textmd{sdim}_q({V}_{4j+1})=\sum _{
h}(-1)^{p(v(h))}q^{h}=\left\{\ba{cc}\frac{q^{2j+1/2}+q^{-2j-1/2}}
{q^{1/2}+q^{-1/2}}, & \mbox{if}\quad
4j+1 \quad \mbox{is odd},\\
\frac{-q^{2j+1/2}+q^{-2j-1/2}} {q^{1/2}+q^{-1/2}}, &\mbox{if}\quad
4j+1 \quad {\mbox{is even}},\ea\right. \eea
 where the sum goes
over all the states labelled by $h$, and we assumed that the
lowest weight vector has $0$ parity.
 Let us note also, that if $\textmd{sdim}_q({V}_{4j+1})=0$, then
it follows $(-q)^{4j+1}=1$. So for the conventional odd
r-dimensional representations, the relation
$\textmd{sdim}_q({V}_{r})=0$ takes place, when $r=N$, $q^N=-1$
(and also $\textmd{sdim}_q({\bar{V}}_{pN})=0$), with odd integers
$N,\; p$. And even dimensional representations have $0$
$q$-superdimension, $\textmd{sdim}_q({V}_{r})=0$
($\textmd{sdim}_q(\bar{V}_{p r})=0$), if $r=2 N $, $q^N=\pm 1$,
and $N, \;p$ are integers.

 It was stated, that in the decompositions of tensor products an
indecomposable representation appears instead of two
representations {\it{only}} if the sum of their
$q$-(super)dimensions is zero (see \cite{PS, S}). The parities of
the lowest weights of $V_{4j_i\!+\!1},\;i=1,2$ in decompositions
differ one from another by $[2(j_2\!-\!j_1)\mathrm{mod}2]$. Taking
this into account, one concludes that the relation
\be
[\textmd{sdim}_q(\bar{V}_{4j_2+1})+\textmd{sdim}_q(V_{4j_1+1})]=0
\label{sdim} \ee
 implies
$\;\;[\left(q^{2(j_2+j_1)+1}+(-1)^{2(j_2+j_1)}\left)\right(q^{2(j_1-j_2)}+
(-1)^{2(j_1-j_2)}\right)=0]$, which is in full agreement with the
equalities (\ref{cc1}, \ref{cc}), with $j=\!j_2,\;
h=\!2j_1\!+\!1$. So, the relation (\ref{sdim}) follows from the
Statement II.

Note, that the definition of the co-product of generator $k$
implies that, if one of the multipliers in the tensor product has
vanishing $q$-superdimension, then the sum of the
$q$-superdimensions over the representations in the decomposition
also is equal to zero.
%
\vspace{-0.5cm}
\paragraph{Remark.} The representation, given by the formulae
(\ref{ind}), is indecomposable in general. However the structure
$\mathcal{I}=\widehat{\bar{V}\oplus\bar{V}}$ having more than two
lowest (highest) weights can be split into the sum of the
indecomposable $\mathcal{I}=\widehat{\bar{V}\oplus {V}}$ ($p=1$ in
(\ref{in})) and the irreducible representations with $0$
$q$-superdimension. It is conditioned by the appropriate values,
which  the coefficients $\beta, \;\gamma $ can acquire in
(\ref{ind}). Such situation happens in the fusions $\otimes^n V_j$
of the irreps due to the nilpotency of the generators $e,\; f$
(see the subsections $5.2$,
 $5.3$).
As we have seen in the discussed examples
$\mathcal{I}^{(12)}_{\{7,5\}}=\widehat{\bar{V}_7\oplus \bar{V}_5}$
splits into two $\mathcal{I}^{(6)}_{\{4,2\}}$-kind representations
at $q^3=-1$ (\ref{333}), but it is not decomposable in
(\ref{i35}). In the following discussion we shall keep the
notation $\mathcal{I}$ for the cases when $p>1$ (\ref{in}) too,
recalling that in the fusions of the irreps they are decomposable.

\subsection{Fusion rules}

Here we intend to derive general fusion rules at roots of unity.
As for the given value of $q$ ($q^N=\pm 1$) the spin
representations are no longer irreps starting from the spin value
$\bar{j}=j_{max}+ \frac 12$ (with $j_{max}$ being the maximal spin
determined by (\ref{jmax1}, \ref{jmax2})), 
 then in the decomposition
 $V_{4j_1+1}\otimes V_{4j_2+1}$, at $(j_1+j_2)\geq
\bar{j}$,
 together with the allowed irreps, also indecomposable representations
 appear.

We can rewrite the formula (\ref{in}) to express the dimensions of
$\mathcal{I}$-representations through the maximal dimension of the
allowed irreps $r_{max}=4 j_{max}+1$.
 \be
\mbox{dim}[\mathcal{I}^{\mathcal{R} p}]=\!\left\{\!\ba{ll}4 N
p=4(2j _{max}+1)p=\!(2
r_{max}+2)p,&q^N=-1,\;\;N\;\;\mbox{even},\;\&\;\;q^N
=1,\;\;N\;\;\mbox{odd},\\ 2N p=2(4j_{max}+1)p=2 r_{max}p,&\quad
\quad \quad q^N=-1, \;\; N\;\; \mbox{is odd
integer}.\ea\right.\label{inj}
 \ee
Here $\mathcal{R}=2N$ or $\mathcal{R}=4N$ denotes the minimal
dimension of $\mathcal{I}$-representations. Note that
$\mathcal{N}=\mathcal{R}/2$.
\paragraph{$\mathbf{V\otimes V}$.}
It is evident from 
(\ref{inj}) that for
$(-q)^\mathcal{N}= 1$ the representations $\mathcal{I}$, which
appear in the tensor product of two arbitrary irreps
$V_{r_1}\otimes V_{r_2}$ when $r_{max}<(r_1+r_2-1)\leq
(2r_{max}-1)$,  can be only with minimal dimensions
($\mathcal{I}^{\mathcal{R} p},\; p=1$), i.e. indecomposable.
In the decomposition (\ref{t}) for the general $q$ the irrep with
maximal dimension is $V_{r_1+r_2-1}$.
For the exceptional values of $q$ the representation
$\bar{V}_{r_1+r_2-1}$ (and hence the remaining $\bar{V}_{r}$,
$r_{max}< r< r_1+r_2-1$) can not turn into the maximal
sub-representation for $\mathcal{I}^{(2 r_{max}p)}$ or
$\mathcal{I}^{2p(r_{max}+1)}$, when $p>1$.

 When $(-q)^\mathcal{N}= 1$
 for representation $V_r$ from the interval $r_{max}< r\leq r_1+r_2-1$, it takes place
 $\alpha_{(\mathcal{R}-r+1)/2}^{(r-1)/4}(q)=0$ (\ref{qnn},
 \ref{qn}), (\ref{inj}). Hence the points of the Statement II
 must be realized for the representations $V_{r}$
 and $V_{\mathcal{R}-r}$. From dimensional analysis it is clear that $r>
\mathcal{R}-r$. In
 agreement with the conclusion of Statement II
%
%
%
%
all the representations $\bar{V}_{r}$, $r>r_{max}$, starting from
$\bar{V}_{r_1+r_2-1}$, are unifying with $V_{\mathcal{R}-r}$ to
produce $\widehat{{\bar{V}}_{r}\oplus
V_{\mathcal{R}-r}}=\mathcal{I}^{( \mathcal{R})}_{\{r,
\mathcal{R}-r\}} $. The other $V_{4j+1}$-s, which do not coincide
with $V_{\mathcal{R}-r}$, survive in this decomposition. By the
solutions to the equation $e\cdot
v'(\frac{\mathcal{R}-r-1}{2})=v_{\frac{r-1}{4}}(\frac{\mathcal{R}-r+1}{2})$,
the states $\{v'(\frac{\mathcal{R}-r-1}{2}),\;f\cdot
v'(\frac{\mathcal{R}-r-1}{2}),...,f^{\mathcal{R}-r-1}\cdot
v'(\frac{\mathcal{R}-r-1}{2})\}$ can be constructed, which
together with the states of representation $\bar{V}_r$ constitute
the indecomposable representation $\mathcal{I}^{(
\mathcal{R})}_{\{r, \mathcal{R}-r\}}$. In conclusion, we have
\bea V_{r_1}\otimes
V_{r_2}=\left(\bigoplus_{r=r_{max}+1}^{r_1+r_2-1} \mathcal{I}^{(
\mathcal{R})}_{\{r,
\mathcal{R}-r\}}\right)\oplus\left(\bigoplus_{r'\leq r_{max},\;
r'\neq\{\mathcal{R}-(r_1+r_2-1),...,\mathcal{R}-r_{max}-1\}}
V_{r'}\right).\label{viv}\eea
This result is in the agreement with the rules derived by another
technique (we are grateful to author of \cite{SB} for the kind
correspondence about this question). About the fusions of the
irreps there was discussion in \cite{Zh}. See also the subsection
$5.5$ for a connection with the case of $sl_q(2)$.

\paragraph{$\mathbf{\otimes^n V}$.}
The tensor product of the finite dimensional representations of
$osp_q(1|2)$ for general $q$ is reduced into a linear combination,
and for $n$ copies of the same representation we can write
 \be
\bigotimes^n V_{4j+1}=\underbrace{V_{4j+1} \otimes V_{4j+1}
\otimes \ldots \otimes
V_{4j+1}}_{n}=\bigoplus_{p=0}^{n/2}\varepsilon^{p j}_{n}V_{4p+1},
\label{VV}\ee
where $\varepsilon^{p j}_{n}$ stands for the multiplicity of the
representation $V_{4p+1}$ and can be calculated in easiest way
using Bratteli diagrams (Fig.\ref{f}) \cite{JimGRAS}. At the left
of $n$-th row of the diagram denomination of tensor product
$\otimes^{n}V_{4j+1}$ is placed, the representations $V_{4p+1}$,
arising in (\ref{VV}) are denoted by dots which are located at the
same row. The representations of the same type $V_{4p+1}$,
regarding to different $n$-s, are arranged in vertical columns.
The multiplicity $\varepsilon^{p j}_{n}$ is determined by the
number of all the possible paths leading from the top of the
diagram to the given representation $V_{4p+1}$ situated at the
level $n$. The paths are formed by the lines connecting the dots.
The intersections of two paths outside of the dots are to be
ignored.
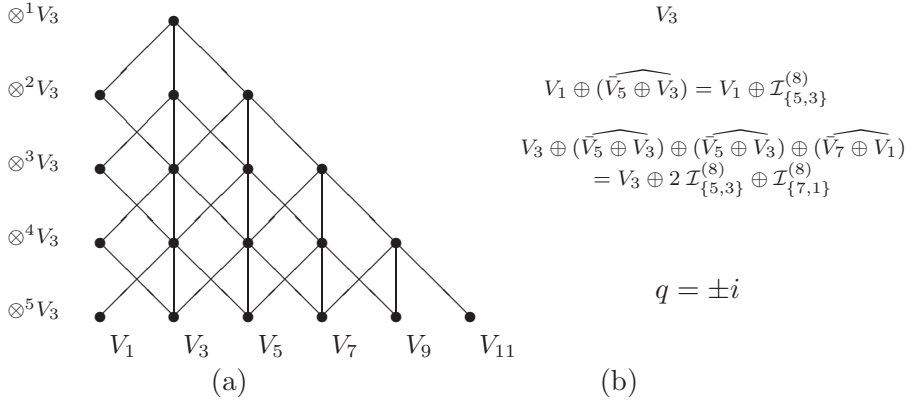
\begin{figure}[t]
\hspace{1 cm} \unitlength=7pt
\begin{picture}(100,20)(17.5,0)

\multiput(24,20)(4,-4){4}{\line(1,-1){4}}
\multiput(24,20)(0,-4){4}{\line(0,-1){4}}
\multiput(24,20)(0,-8){2}{\line(-1,-1){4}}
\multiput(28,16)(0,-4){3}{\line(0,-1){4}}
\multiput(32,12)(0,-4){2}{\line(0,-1){4}}
\put(36,8){\line(0,-1){4}}
\multiput(20,16)(0,-8){2}{\line(1,-1){4}}
\put(24,12){\line(1,1){4}}
\multiput(32,12)(-4,-4){2}{\line(-1,-1){4}}
\multiput(24,16)(4,-4){3}{\line(-1,-1){4}}
\multiput(24,16)(4,-4){3}{\line(1,-1){4}}
\multiput(20,12)(4,-4){2}{\line(1,-1){4}}
\multiput(36,8)(-4,-4){1}{\line(-1,-1){4}}
\multiput(24,12)(4,-4){2}{\line(1,-1){4}}
\put(24,8){\line(-1,-1){4}}

\put(15,20){\scriptsize $\otimes^1 V_{3}$}
\put(15,16){\scriptsize$\otimes^2 V_{3}$} \put(15,12){\scriptsize
$\otimes^3 V_{3}$}\put(15,8){\scriptsize $\otimes^4 V_{3}$}
\put(15,4){\scriptsize $\otimes^5 V_{3}$}

\put(20.5,2){\small $V_{1}$}\put(24.5,2){\small
$V_{3}$}\put(28.5,2){\small $V_{5}$}\put(32.5,2){\small
$V_{7}$}\put(36.5,2){\small $V_{9}$}\put(40.5,2){\small $V_{11}$}

 \put(50,20){\scriptsize
$V_{3}$} \put(44,16){\scriptsize$V_1 \oplus
(\widehat{\bar{V}_{5}\oplus V_{3}})=V_1\oplus
\mathcal{I}^{(8)}_{\{5,3\}}$} \put(42,12){\scriptsize $\ba{c}V_{3}
\oplus (\widehat{\bar{V}_{5}\oplus V_{3}})\oplus
(\widehat{\bar{V}_{5}\oplus V_{3}})\oplus
(\widehat{\bar{V}_{7}\oplus V_1})\\=V_{3}\oplus
2\;\mathcal{I}^{(8)}_{\{5,3\}}\oplus
\mathcal{I}^{(8)}_{\{7,1\}}\ea $}
\put(50,5){$q=\pm i$}

\multiput(24,20)(0,-4){5}{\circle*{0.5}}
\multiput(20,16)(0,-4){4}{\circle*{0.5}}
\multiput(28,16)(0,-4){3}{\circle*{0.5}}
\multiput(32,12)(0,-4){3}{\circle*{0.5}}
\multiput(36,8)(0,-4){2}{\circle*{0.5}}
\put(40,4){\circle*{0.5}}\put(28,4){\circle*{0.5}}

\put(26,0){(a)} \put(47,0){(b)}
\end{picture}
 \caption{Bratteli diagram for the
spin-$\frac{1}{2}$ irreps at general $q$ (a), fusion
 for $q^2=-1$ (b).}\label{f}
\end{figure}

When $q$ is a root of the unity, the decomposition (\ref{VV})
remains unchanged while $n j\leq 2 j_{max}$. We are interested in
the representations emerging in the fusions of the fundamental
irreps, i.e $j=1/2$. The minimal $n$ for which indecomposable
representations appear in the decomposition of $\otimes^{n}V_{3}$,
is $n\!=\!2\bar{j}\equiv\! 2 j_{max}\!+\!1$. The fusions
corresponding to two possibilities (\ref{qnn}) and (\ref{qn}) are
respectively:
\bea\label{vqnn}\otimes^{2\bar{j}}V_{3}&=&\oplus_{j<\bar{j}}
(\varepsilon^{j\;\;3}_{2\bar{j}}-\delta_{j,\bar{j}-1/2})V_{4j+1}+\mathcal{I}^{(4N)}_{\{2N+1,2N-1\}},\\\label{vqn}
\otimes^{2\bar{j}}V_{3}&=&\oplus_{j<\bar{j}}
(\varepsilon^{j\;\;3}_{2\bar{j}}-\delta_{j,\bar{j}-1})V_{4j+1}+\mathcal{I}^{(2N)}_{\{N+2,N-2\}}.
\eea
%
The associativity of the tensor product allows to obtain this
formula, using (\ref{viv}) for $V_{4j_{max}+1}\otimes V_3$.
%
From (\ref{jmax1}, \ref{jmax2}) it follows
%
%
$\bar{j}=N/2$ for the first case (\ref{jmax1}) and
$\bar{j}=\frac{N+1}{4}$ for the second case (\ref{jmax2}). As in
the decomposition of $\otimes^{ 2\bar{j}}V_{3}$ only the
representation with maximal dimension $V_{4\bar {j}+1}$ becomes
$\bar{V}_{4\bar{j}+1}$, then to reveal the structure of the
possible indecomposable representations (with maximal proper
sub-representation $\bar{V}_{4\bar{j}+1}$), which can appear in
agreement with the Statement II, one has to check invariant
sub-representations of $\bar{V}_{4\bar{j}+1}$. And one can verify
that the relations $\alpha_{2\bar{j}}^{\bar{j}}(q)=0$
(\ref{jmax1}) and $\alpha_{2\bar{j}-1}^{\bar{j}}(q)=0$
(\ref{jmax2}) take place, and the proper sub-representation of
$\bar{V}_{4\bar{j}+1}$ is the representation spin-$(\bar{j}-1/2)$
or spin-$(\bar{j}-1)$ for the cases (\ref{jmax1}) or (\ref{jmax2})
correspondingly. 
%
%
In the fusions the invariant sub-space of the representation
$\bar{V}_{4 \bar{j}+1}$ becomes linearly dependent with
representation space of $V_{4\bar{j}-1}$ for the case
(\ref{jmax1}) (correspondingly with $V_{4\bar{j}-3}$, for the case
(\ref{jmax2})), and then $\bar{V}_{4 \bar{j}+1}$ together with
other $4\bar{j}-1$ vectors (with $4\bar{j}-3$ vectors),
forms new $(4\bar{j}+1)+(4\bar{j}-1)=4N$ dimensional
indecomposable representation $\mathcal{I}^{(4N)}_{\{2N+1,2N-1\}}$
(\ref{vqnn}) ($(4\bar{j}+1)+(4\bar{j}-3)=2N$ dimensional
indecomposable representation $\mathcal{I}^{(2N)}_{\{N+2,N-2\}}$
(\ref{vqn})). As the multiplicity of $V_{4\bar{j}+1}$ in the
fusion is one, then in (\ref{vqnn}, \ref{vqn}) the number of the
indecomposable representations is also equal to one.
The multiplicities $\varepsilon^{j\;\;3}_{2\bar{j}}$ can be
checked by means of Bratteli diagrams, as in case of general $q$.

 Now, let us present a scheme for
derivation of fusion $\otimes^{n}V_{3}$ for an arbitrary $n$.
To determine the decomposition of tensor product for the
exceptional values of $q$, using (\ref{VV})
(defined for the general $q$), the following scheme can work:
if $(nj)>j_{max}$, the highest-dimensional representation
${V}_{4nj+1}$ (appears in (\ref{VV}) with multiplicity
$\varepsilon^{(nj)\;j}_{n}=1$) turns to be $\bar{V}_{4nj+1}$. If
 $\textmd{sdim}_q(\bar{V}_{4nj+1})=0$, this representation
splits into the direct sum of the irreps $V_r$, with
$\textmd{sdim}_q({V}_{r})=0$. It follows from the values of the
enlarged center elements $e^{\mathcal{N}}=0,\; f^{\mathcal{N}}=0$
and the dimensional analysis before the formula (\ref{sdim}).
Otherwise, if the largest invariant sub-representation of
$\bar{V}_{4nj+1}$ is a ($4s+1$)-dimensional representation (i.e.
$s$ is the maximal $h/2$ for which $\alpha^{n j}_{h+1}(q)=0$ in
(\ref{alpha})), then the
largest $\mathcal{I}$-representation %
arises from unification of $\bar{V}_{4nj+1}$ with one of
$V_{4s+1}$ appearing in decomposition:
$\widehat{\bar{V}_{4nj+1}\oplus\tilde {V}_{4s+1}}=\mathcal{I}^
{(4(nj+s)+2)}_{\{4nj+1,4s+1\}}$, where
$\tilde{V}_{4s+1}=\left\{\ba{cc}\bar V_{4s+1} \;\;{\rm{if}}\;\;s>j_{max}\\
V_{4s+1} \;\;{\rm{if}}\;\; s\leq j_{max}\ea\right.$. Then one must
consider in the same way the representation next to the
highest-dimensional, if it is not an irrep, i.e.
$\bar{V}_{4nj-1}$, taking into account its multiplicity, which can
be reduced by one, if $s=nj-\frac12$, and so on.

As an example let us consider degeneracy of $\otimes^3 V_{3}=V_{1}
\oplus 3 V_{3}\oplus 2 V_{5}\oplus V_{7}$ at $q=\pm i$ (see figure
2). As $\alpha^{3/2}_{1}(\pm i)=0$, then $\bar{V}_{7}\supset {V}
_{1}$, so there is
$\mathcal{I}^{(8)}_{\{7,1\}}=\widehat{\bar{V}_{7}\oplus {V}_{1}}
$. Then $\bar{V}_{5}\supset V_{3}$, as $\alpha^{1}_{2}(\pm i)=0$,
and two $\bar{V}_{5}$ become the part of two
$\mathcal{I}^{(8)}_{\{5,3\}}=\widehat{\bar{V}_{5}\oplus V_{3}}$.
So we have $V_{3}\oplus 2 \mathcal{I}^{(8)}_{\{5,3\}}\oplus
\mathcal{I}^{(8)}_{\{7,1\}}$. For arbitrary $n$ at $q=\pm i$
moving in the same way, the relation (\ref{vvk}) can be traced,
finding multiplicities from the dimensional analysis. For
$\otimes^4 V_3$, at $q^3=-1$, the analysis gives
${\mathrm{sdim}}_q( V_3)=0,\; {\mathrm{sdim}}_q (\bar{V}_9)=0$, as
$e^3=0,\; f^3=0$. Hence $\bar{V}_9=\oplus^3 V_3$ and $\otimes^4
V_3=9 V_{3}\oplus 3(\mathcal{I}^{(12)}_{7,5})\oplus
3(\mathcal{I}^{(6)}_{5,1})=9 V_{3}\oplus
6(\mathcal{I}^{(6)}_{4,2})\oplus 3(\mathcal{I}^{(6)}_{5,1})$.
%
%
%

As it was noted the representation $\mathcal{I}^{(p
\mathcal{R})}=\widehat{\bar{V}_{r}\oplus\bar{V}_{p
\mathcal{R}-r}}, \;\; p>1$, arising in the fusions $\otimes^n
V_3$, is decomposable. 
 The
$\mathcal{N}$-nilpotency of $e,\; f$
($
\mathcal{N}=\mathcal{R}/2$) helps us to determine how many and
what kind of irreducible invariant sub-representations has the
proper sub-representation $\bar{V}_{r}$ of $\mathcal{I}^{(p
\mathcal{R})}_{\{r,p\mathcal{R}-r\}}$. One can directly count that
$\bar{V}_{r}$ contains $p$ proper irreducible subspaces with
dimension $\mathbf{r_p}$,
\be \label{rp} p=[r/{\mathcal{N}}],\qquad
\mathbf{r_p}={\mathcal{N}}-r+{\mathcal{N}} p,\ee
 (here $[x]$ denotes the integer part of $x$).
So,  $\bar{V}_{r}\supset \underbrace{V_{\mathbf{r_p}}\oplus
V_{\mathbf{r_p}}\cdots\oplus V_{\mathbf{r_p}}}_p$,
$\mathbf{r_p}\leq 4j_{max}+1$.
The highest weight vectors of $\bar{V}_{r}$, together with
$v_j(j),\; j=(r-1)/4)$, now are $v_j(2 j^{\;i}),\; 2 j^{\;i}=-2
j+i \mathcal{R}/2-1\equiv(i \mathcal{R}-r-1 )/2$, $i=1,...,p$.
 The linear dependence can be
established between $p\; \mathbf{r_p}$ vectors
belonging to the mentioned proper irreducible subspaces of
$\bar{V}_{r}$ and the corresponding vectors with the same weights
of the representation $\bar{V}_{p\mathcal{R}-r}$. All these
vectors have null norms. And new, not orthogonal to them vectors
$v'(h)$ are to be constructed similarly to the case when $p=1$.
So, the internal structure of $\mathcal{I}^{(p
\mathcal{R})}_{\{r,p\mathcal{R}-r\}}$ is characterized  as
 \bea
 \mathcal{I}^{(p
\mathcal{R})}_{r,p\mathcal{R}-r}=\underbrace{\mathcal{I}^{(
\mathcal{R})}_{\{\mathcal{R}-\mathbf{r_p},\mathbf{r_p}\}}\oplus
\mathcal{I}^{(
\mathcal{R})}_{\{\mathcal{R}-\mathbf{r_p},\mathbf{r_p}\}}\cdots
\oplus \mathcal{I}^{(
\mathcal{R})}_{\{\mathcal{R}-\mathbf{r_p},\mathbf{r_p}\}}}_p.\label{ip}
\eea
 The sign "$=$" in (\ref{ip}) means an isomorphism, as the
eigenvalues of the generator $k$  on the states in  r.h.s differ
by common multipliers from the ones of
$\mathcal{I}^{(\mathcal{R})}_{\{\mathcal{R}-\mathbf{r_p},\mathbf{r_p}\}}=
\widehat{\bar{V}_{\mathcal{R}-\mathbf{r_p}}\oplus
V_{\mathbf{r_p}}}$.

 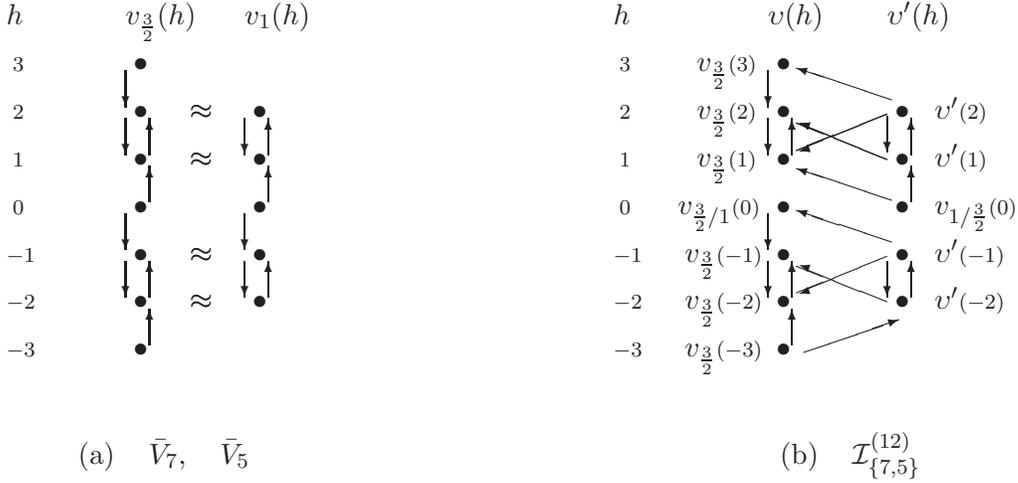
\begin{figure}[h]
 \hspace{1 cm} \unitlength=9pt
 \begin{picture}(100,25)(15,0)

\put(20,22){$v_{\frac{3}{2}}(h)$} \put(25,22){$v_{1}(h)$}
 \multiput(20,20)(0,-2){7}{$\;\bullet$}
 \multiput(25,18)(0,-2){5}{$\;\bullet$}
 \put(15,22){$h$}%
 \put(15,20){\scriptsize$\;3$}
 \put(15,18){\scriptsize $\;2$} \put(15,16){\scriptsize $\;1$}
 \put(15,14){\scriptsize $\;0$} \put(15,12){\scriptsize
 $-1$} \put(15,10){\scriptsize $-2$}
 \put(15,8){\scriptsize $-3$}

 \multiput(26,10.5)(0,4){2}{\vector(0,1){1.5}}
 \multiput(25,18)(0,-4){2}{\vector(0,-1){1.5}}
\put(25,12){\vector(0,-1){1.5}}\put(26,16.5){\vector(0,1){1.5}}
 \multiput(21,14.5)(0,2){2}{\vector(0,1){1.5}}
 \multiput(20,20)(0,-2){2}{\vector(0,-1){1.5}}
\multiput(21,8.5)(0,2){2}{\vector(0,1){1.5}}
\multiput(20,14)(0,-2){2}{\vector(0,-1){1.5}}

\multiput(22.7,18)(0,-2){2}{$\approx$}
\multiput(22.7,12)(0,-2){2}{$\approx$}

 \put(18,3.5){(a) $\;\;\;\bar{V}_{7},\quad \bar{V}_5$}


\put(47,22){$\upsilon(h)$} \put(52,22){$\upsilon'(h)$}
 \multiput(47,20)(0,-2){7}{$\;\bullet$}
 \multiput(52,18)(0,-2){5}{$\;\bullet$}
 \put(40.5,22){$h$}%
 \put(40.5,20){\scriptsize$\;3$}
 \put(40.5,18){\scriptsize $\;2$} \put(40.5,16){\scriptsize $\;1$}
 \put(40.5,14){\scriptsize $\;0$} \put(40.5,12){\scriptsize
 $-1$} \put(40.5,10){\scriptsize $-2$}
 \put(40.5,8){\scriptsize $-3$}
\put(44,20){$v_{\frac{3}{2}}$\scriptsize${(3)}$}
\put(44,18){$v_{\frac{3}{2}}$\scriptsize${(2)}$}
\put(44,16){$v_{\frac{3}{2}}$\scriptsize${(1)}$}
\put(43.2,14){$v_{\frac{3}{2}/1}${\scriptsize${(0)}$}}
\put(43.5,12){$v_{\frac{3}{2}}$\scriptsize${(-1)}$}
\put(43.5,10){$v_{\frac{3}{2}}$\scriptsize${(-2)}$}
\put(43.5,8){$v_{\frac{3}{2}}$\scriptsize${(-3)}$}

\put(54,18){$\upsilon'$\scriptsize$(2)$}
\put(54,16){$\upsilon'$\scriptsize$(1)$}
\put(54,14){$v_{1/\frac{3}{2}}${\scriptsize${(0)}$}}
\put(54,12){$\upsilon'$\scriptsize$(-1)$}
\put(54,10){$\upsilon'$\scriptsize$(-2)$}

 \multiput(53,10.5)(0,4){2}{\vector(0,1){1.5}}
 \multiput(52,18.1)(0,-4){1}{\vector(0,-1){1.5}}
\put(52,12){\vector(0,-1){1.5}}\put(53,16.5){\vector(0,1){1.5}}
 \multiput(48,16.5)(0,2){1}{\vector(0,1){1.5}}
 \multiput(47,20)(0,-2){2}{\vector(0,-1){1.5}}
\multiput(48,8.5)(0,2){2}{\vector(0,1){1.5}}
\multiput(47,14)(0,-2){2}{\vector(0,-1){1.5}}
\put(52.2,18.8){\vector(-3,1){4}}
\put(52.2,14.6){\vector(-3,1){4}}
\put(52.2,12.8){\vector(-3,1){4}} \put(48.5,8.2){\vector(3,1){4}}
\multiput(52,18.2)(0,-6){2}{\line(-5,-2){3.8}}
\multiput(52,16.3)(0,-6){2}{\line(-5,2){3.8}}
\multiput(48.5,16.7)(0,-6){2}{\vector(-3,-1){0.2}}
\multiput(48.5,17.78)(0,-6){2}{\vector(-3,1){0.2}}

\put(52,18.2){\line(-5,-2){3.8}}
\put(52,16.3){\line(-5,2){3.8}}
\put(48.5,16.7){\vector(-3,-1){0.2}}
\put(48.5,17.78){\vector(-3,1){0.2}}

\multiput(22.7,18)(0,-2){2}{$\approx$}
\multiput(22.7,12)(0,-2){2}{$\approx$}

 \put(47.5,3.5){(b) $\;\;\;\mathcal{I}^{(12)}_{\{7,5\}}$}
 \end{picture}
 \vspace{-1.5cm}
 \caption{Representations  $\bar{V}_{7},\; \bar{V}_5$ (a)
  and
 $\mathcal{I}^{(12)}_{\{7,5\}}$ (b) in $\otimes^3 V_3$, $q^3=-1$.}\label{fr1}

 \end{figure}

For illustration we are representing the emergence of
$\mathcal{I}^{(12)}_{\{7,5\}}$ at $q^3=-1$ in the decomposition of
$V_3\otimes V_3 \otimes V_3$. In the Fig. (\ref{fr1} a)  two non
completely reducible representations $\bar{V}_{7},\; \bar{V}_5$
are shown, which have two proper 2-dimensional subspaces with
$h=1,\;2$ and $h=-1,\;-2$. There is a linear dependence between
the corresponding vectors $v_{\frac{3}{2}}(h)$ and $v_{1}(h)$,
which in the figure has denoted by the symbol "$\approx$". In the
Fig. (\ref{fr1} b) the structure of
$\mathcal{I}^{(12)}_{\{7,5\}}=\widehat{\bar{V}_{7}\oplus
\bar{V}_{5}}$ is presented. This representation is decomposed into
two indecomposable representations
$\mathcal{I}^{(6)}=\widehat{\bar{V}_{4}\oplus V_{2}}$. By means of
$v_{\frac{3}{2}/1}(0)$ and $v_{1/\frac{3}{2}}(0)$ two mutually
orthogonal vectors are denoted, which are some linear
superpositions of the vectors $v_{\frac{3}{2}}(0)$ and $v_{1}(0)$.

 As
an another example let us observe the case $q^2=-1$, which will
give the exact decomposition of (\ref{vvk}). $\mathcal{I}^{(p 8)}$
can be composed by $\widehat{\bar{V}_{4p+1}\oplus\bar{V}_{4p-1}}$
or $\widehat{\bar{V}_{4p+3}\oplus\bar{V}_{4p-3}}$ (consideration
of the even dimensional irreps would enlarge the possibilities by
$\widehat{\bar{V}_{4p+2}\oplus\bar{V}_{4p-2}}$). It can be checked
straightly (\ref{rp}), that $\bar{V}_{4p+1}$ ($\bar{V}_{4p+3}$)
has $V_3$ ($V_1$)-type invariant sub-irreps. It gives
\be \mathcal{I}^{(8 p)}_{\{4p+1,4p-1\}}=
\underbrace{\mathcal{I}^{(8)}_{\{5,3\}}\oplus
\mathcal{I}^{(8)}_{\{5,3\}}\oplus\cdots
\mathcal{I}^{(8)}_{\{5,3\}}}_p,\quad\mathcal{I}^{(8
p)}_{\{4p+3,4p-3\}}= \underbrace{\mathcal{I}^{(8)}_{\{7,1\}}\oplus
\mathcal{I}^{(8)}_{\{7,1\}}\oplus\cdots
\mathcal{I}^{(8)}_{\{7,1\}}}_p.\ee
%
{$\mathbf{V\otimes \mathcal{I}}$, $\mathbf{\mathcal{I}\otimes
\mathcal{I}}$.
} The fusion rules of the products like $V\otimes \mathcal{I}$ and
$\mathcal{I}\otimes \mathcal{I}$  can be found either from the
decomposition of $\otimes^n V$, recalling the associativity
property of the product (quite analogous to the cases (\ref{333i},
\ref{i8v}, \ref{ii})), or in this way: let
$\mathcal{I}=\widehat{\bar{V'}\oplus\tilde{V''}}$, then one must
write down the tensor product for $V'\oplus V''$ at general $q$,
and analyze it's deformation at the exceptional values quite
similar to the case $\otimes^n V$. Let us present the
decomposition of
$\mathcal{I}^{(\mathcal{R})}_{\{r,\mathcal{R}-r\}}\otimes
V_3=(\widehat{\bar{V}_{r}\oplus V}_{\mathcal{R}-r})\otimes V_3$.
From the fusion rule at general $q$ (we assume that
$\mathcal{R}-r>1$)
\bea (V_{r}\oplus V_{\mathcal{R}-r})\otimes V_3=V_{r+2}\oplus
V_{r} \oplus V_{r-2}\oplus V_{\mathcal{R}-r+2}\oplus
V_{\mathcal{R}-r}\oplus V_{\mathcal{R}-r-2}, \eea
such decomposition at roots of unity ($(-q)^{\mathcal{R}/2}=1$,
see (\ref{inj})) will be followed:
\bea \nn \mathcal{I}^{(\mathcal{R})}_{\{r,\mathcal{R}-r\}}\otimes
V_3=\widehat{\bar{V}_{r+2}\oplus V}_{\mathcal{R}-r-2}\oplus
\widehat{\bar{V}_{r}\oplus V}_{\mathcal{R}-r} \oplus
\widehat{\tilde{V}_{r-2}\oplus \tilde{V}}_{\mathcal{R}-r+2}=\\
\mathcal{I}^{(\mathcal{R})}_{\{r+2,\mathcal{R}-r-2\}}\oplus
\mathcal{I}^{(\mathcal{R})}_{\{r,\mathcal{R}-r\}}\oplus \left\{
\ba{cc} V_{r_{max}}\oplus V_{r_{max}}, &\;\;\mathrm{if} \quad r-2=
r_{max},\quad \mathcal{R}/2 \; \;\mathrm{odd}\\
\mathcal{I}^{(\mathcal{R})}_{\{r_{max}+2,r_{max}\}},&\;\;
\mathrm{if}
\quad r-2= r_{max},\quad \mathcal{R}/2 \;\;\rm{even}\\
\mathcal{I}^{(\mathcal{R})}_{\{r-2,\mathcal{R}-r+2\}},& \;\;
\mathrm{if} \quad r-2 > r_{max}\;.\ea
 \right.
 \eea
%
%
 \subsection{
 Clebsh-Gordan coefficients 
 }

\paragraph{ Clebsh-Gordan coefficients.}
To make sure by \textit{direct constructions} that in the
decomposition of the tensor products two representations at
$q^N=\pm 1$ are unified in the manner described above (Statement
II), $\widehat{\bar{V}_{4j+1}\oplus{V} _{4s+1}}$, one can
calculate Clebsh-Gordan coefficients for the representations of
this algebra and check the linear dependence of the vectors
belonging to $\bar{V}_{4j+1}$ and ${V}_{4s+1}$.

Let us remind definition of Clebsh-Gordan (CG) coefficients: if
$V_{4j+1}=\{v_j(h)\},\; h=-2j,...,2j$, is an irrep arising in the
decomposition (\ref{v*v}), then it's states are defined as
 \bea
v_j(h)=\!\!\!\sum_{h_1+h_2=h}\!\!\!\!\!\!C\left(^{j_1,j_2,j}_{h_1,h_2,
h}\right)v_{j_1}(h_1)\otimes v_{j_2}(h_2). \label{vjh}
 \eea

The second point of the Statement II affirms that in the fusions
two representations ${V}_{4j+1},\;\;{V}_{4s+1}$, $j>s$ are
replaced by an indecomposable one, when ${V}_{4s+1}$ and a
sub-representation of $V_{4j+1}$, $\{v_j(h) \}$ with $h=-2s,\cdots
2s$, are linearly dependent, i.e
$\{v_j(-2s),\ldots,v_j(2s)\}\approx\{v_s(-2s),\ldots ,v_s(2s)\}$.
In terms of Clebsh-Gordan coefficients it means
 \be
\{C(^{j_1,j_2,j}_{h_1,h_2,h})\}_{h_1+h_2=h}\approx
\{C(^{j_1,j_2,s}_{h_1,h_2,h})\}_{h_1+h_2=h},\quad h=-2s,\ldots
2s.\label{cg}
 \ee
This relation implies that the functions
$C(^{j_1,j_2,j}_{h_1,h_2,h})/C(^{j_1,j_2,s}_{h_1,h_2,h})$ do not
depend on the variables $h_1,\;h_2$.

Here we calculate the coefficients up to the normalization
factors, which are inessential when $q$ is given by a root of
unity. Using the highest weight method \cite{KS} and the
co-product (\ref{coprod}) for the representations (\ref{odd}) we
find the following expressions for $C(^{j_1,j_2,j}_{h_1,h_2,h})$,
$\;\;\;h=2j$,
 \bea
C(^{j_1,j_2,j}_{h_1,2j-h_1,2j})=\prod_{g=h_1+1}^{2j_1}\left(
\frac{(-1)^{p_{j_1,g+1}}q^{2j-g+1}\beta_{2j-g}^{j_2}(q)}{\beta_{g-1}^{j_1}(q)}\right)
C(^{j_1,j_2,j}_{2j_1,2j-2j_1,2j}). \label{cjj}\eea
The parity $p_{j_1,h_1}$ of the state $v_{j_1}(h_1)$ can be
determined in this way: if the lowest weight vectors in the r.h.s
of (\ref{vjh}) have even parity, then
$(-1)^{p_{j_1,h_1}}=(-1)^{2j_1+h_1}$.
Acting by the operator $f^{2j-h}$ on the both sides of the
equation (\ref{vjh}), where $v_j(2j)$ stands on the l.h.s., we
arrive at
 \bea
C(^{j_1,j_2,j}_{h_1,h_2,h})=\prod_{g=h+1}^{2j}(\gamma_{g}^{j}(q))
^{-1} \sum_{r=0}^{2j-h}[^{2j-h}_{\;\;\;r}]_{-q^{-1}}(-1)^{r
p_{j_1,h'_1}}q^{r(h'_1)}\prod_{r_1=h'_1}^{h_1+1}\gamma_{r_1}^
{j_1}(q)\prod_{r_2=h'_2}^{h_2+1}\gamma_{r_2}^{j_2}(q)
\;C(^{j_1,j_2,j}_ {h'_1,h'_2,2j}), \label{cgc}
 \eea
where $h'_1=h_1+(2j-h-r),\;\;h'_2=2j-h'_1=h_2+r$, and
$p_{j_1,h'_1}$ is the parity of the state $v_{j_1}(h'_1)$. The
$q$-binomial coefficients are defined by the formula (\ref{[]}).
We fix the coefficients up to a normalization constant, as for $q$
being a root of unity, their ratios become important rather than
coefficients themselves.

Note, that in different works there are computed formulas for
Clebsh-Gordan coefficients with fixed quantities $\beta, \gamma$
\cite{S, cg},  particularly in the work \cite{S} a specific case
$j_2=1/2$, $C(^{j_1,1/2,j}_{h_1,h_2,h})$ is given, which coincides
with our computations up to the normalization factors.

If one checks the relations (\ref{cg}) 
directly using the formulas (\ref{cgc}), then one has to remove
all the possible common zeroes and singularities appearing in the
coefficients of the vectors $v_j(h)$ ($v_s(h)$) (\ref{vjh}) at the
corresponding exceptional values of $q$ and to verify that the
ratios
$C(^{j_1,j_2,j}_{h_1,h_2,h})/C(^{j_1,j_2,j}_{{\tilde{h}}_1,{\tilde{h}}_2,h})$
coincide with
$C(^{j_1,j_2,s}_{h_1,h_2,h})/C(^{j_1,j_2,s}_{{\tilde{h}}_1,{\tilde{h}}_2,h})$,
$h_1+h_2={\tilde{h}}_1+{\tilde{h}}_2\equiv h \in{-2s,\cdots,2s}$.
The quantities $\beta_h^j$, $\gamma_ h^j$ can be defined as
$\beta_h^j=1,\;\; \gamma_h^j=\alpha_h^j$ for the allowed irreps.
And one must take into account that $\alpha_h^j(q)=\alpha_h^s(q)$,
which follows from $(-q)^{2j+2s+1}=1$ (\ref{alpha}). {{See
Appendix 
for the case of $V_{j_{max}}\otimes V_3$.

On the other hand when $j>j_{max}$ and $\alpha_{2s+1}^j(q)=0$,
i.e. $(-q)^{2s+2j+1}=1$, then
the vector $ v_j(2s)
$ is also
a highest weight vector
. 
 Hence the highest weight method can be applied
also for this vector to find the ratios of it's CG coefficients.
And we can write formulas similar to (\ref{cjj}, \ref{cgc}),
replacing $2j$ by $2s$ (below we suppose $s\geq j_1$)
\bea\label{cjs}
C(^{j_1,j_2,j}_{h_1,2j-h_1,2s})=\prod_{g=h_1+1}^{2j_1}\left(
\frac{(-1)^{p_{j_1,g+1}}q^{2s-g+1}\beta_{2s-g}^{j_2}}{\beta_{g-1}^{j_1}}\right)
C(^{j_1,j_2,j}_{2j_1,2s-2j_1,2s}),\eea
\bea
C(^{j_1,j_2,j}_{h_1,h_2,h})=\prod_{g=h+1}^{2s}(\gamma_{g}^{j}(q))
^{-1} \sum_{r=0}^{2s-h}[^{2s-h}_{\;\;\;r}]_{-q^{-1}}(-1)^{r
p_{j_1,h'_1}}q^{r(h'_1)}\prod_{r_1=h'_1}^{h_1+1}\gamma_{r_1}^
{j_1}(q)\prod_{r_2=h'_2}^{h_2+1}\gamma_{r_2}^{j_2}(q)
\;C(^{j_1,j_2,j}_ {h'_1,h'_2,2s}), \label{cjc}
\eea
 where now $h'-1=h_1+(2s-h-r),\;\;h'_2=2s-h'_1$, $-2s \leq h <2s$.
Comparing these expressions with $C(^{j_1,j_2,s}_{h_1,2j-h_1,h})$,
obtained from the formulas (\ref{cjj}, \ref{cgc}) 
, we see that it ensures the validity of the relations (\ref{cg}):
$\frac{C(^{j_1,j_2,j}_{h_1,2j-h_1,2s})}{C(^{j_1,j_2,s}_{h_1,2j-h_1,2s})}
=
\frac{C(^{j_1,j_2,j}_{2j_1,2s-2j_1,2s})}{C(^{j_1,j_2,s}_{2j_1,2s-2j_1,2s})}$,
$\frac{C(^{j_1,j_2,j}_{h_1,2j-h_1,h})}{C(^{j_1,j_2,s}_{h_1,2j-h_1,h})}
={{\prod}}_{g=h+1}^{2s}\frac{\gamma_{g}^{s}(q)}{\gamma_{g}^{j}(q)}
\frac{C(^{j_1,j_2,j}_{2j_1,2s-2j_1,2s})}{C(^{j_1,j_2,s}_{2j_1,2s-2j_1,2s})}$,
$-2s \leq h <2s$.

%
%
}}
  In the decomposition (\ref{viv}) for the irreps $V_{r'}$ 
the coefficients of the expansion (\ref{vjh}) are to be obtained
just from the formulas (\ref{cgc}), 
fixing the 
values of $q$. For the indecomposable representations
$\mathcal{I}^{(\mathcal{R})}_{\{r,\mathcal{R}-r\}}=\{v_{\mathcal{I}}(-2j),
\cdots, v_{\mathcal{I}}(2j); v'_{\mathcal{I}}(-2s),\cdots,
v'_{\mathcal{I}}(2s)\}$, $r=4j+1,\;\mathcal{R}\!-r=4s+1$,
$(-q)^{2j+\!2s\!+\!1}\!=\!1$
\bea v_{\mathcal{I}}(h)=\sum_{h_1+h_2=h}\!\!\!
C_{\mathcal{I}}\left(^{j_1,j_2,j}_{h_1,h_2,
h}\right)v_{j_1}(h_1)\otimes v_{j_2}(h_2),\quad
v'_{\mathcal{I}}(h)=\sum_{h_1+h_2=h}\!\!\!
C'_{\mathcal{I}}\left(^{j_1,j_2,s}_{h_1,h_2,
h}\right)v_{j_1}(h_1)\otimes v_{j_2}(h_2), \eea
 the
coefficients $C_{\mathcal{I}}\left(^{j_1,j_2,j}_{h_1,h_2,
h}\right)$ for the vectors $ v_{\mathcal{I}}(h)
$ also can be calculated from 
(\ref{cgc}) in the limit when $q$ is a root of unity (only one
must be careful, as now for the values $h<-2s$ there are common
overall factors like $[2j+2s+1]_{(-q)^{1/2}}$ (\ref{[]}) which are
 cancelled by choice $\gamma_{2s+1}^j(q)=\alpha_{2s+1}^j(q)$). As
for the vectors $
v'_{\mathcal{I}}(h)
$, here the
coefficients can be obtained 
using the relation $e \cdot
{v'}_{\mathcal{I}}(2s)={\tilde{\beta}}^s_{2s}(q)v_{\mathcal{I}}(2s+1)$
(\ref{ind}). The resulting expression for the coefficients of
$v'_{\mathcal{I}}(2s)$ is
$C'_{\mathcal{I}}\left(^{j_1,j_2,s}_{h_1,h_2,2s}\right)=$
\bea
\!=\!\! \prod_{i=h_1}^{2j_1-1}\! f_s(j_1,j_2,i)
C'_{\mathcal{I}}\left(\!^{j_1,j_2,s}_{2j_1,2s-2j_1,2s}\!\right)\!+\!\!
\frac{{\tilde{\beta}}^s_{2s}(q)}{f_s(j_1,j_2,h_1)}\sum_{i'=h_1}^{2j_1-1}\!
\prod_{\; i=h_1}^{i'}f_s(j_1,j_2,i)\!
\frac{q^{2s-i'}}{{\beta}^{j_1}_{i'}(q)}
C_{\mathcal{I}}\left(\!^{j_1,j_2,j}_{i'+1,2s-i',2s+1}\!\right)\!,\label{cih}\eea
where
$f_s(j_1,j_2,i)=(-1)^{p_{j_1,i+2}}q^{2s-i}\frac{\beta^{j_2}_{2s-i-1}(q)}
{\beta^{j_1}_{i}(q)}$ is a function entering into the formulas
(\ref{cjj}, \ref{cjs}). Hence the first summand corresponds to the
solution of the homogeneous equation. The indefiniteness of that
term (coming from
$C'_{\mathcal{I}}\left(\!^{j_1,j_2,s}_{2j_1,2s-2j_1,2s}\!\right)$)
can be absorbed by redefinition of the vectors $v'(h)$, which are
determined up to a solution to the homogeneous equation $e \cdot
v(h)=0$. 
Acting by the operator $f^{2s-h}$ on $v'(2s)$ it is possible to
find out the coefficients corresponding to the remaining vectors
$v'_{\mathcal{I}}(h)$. We obtain the following expression for
$C'_{\mathcal{I}}$ (below $h'_1=h_1+(2s-h-r),\;\;h'_2=2s-h'_1$)
%
\bea
C'_{\mathcal{I}}\!(^{j_1,j_2,s}_{h_1,h_2,h})=\!\prod_{g=h+1}^{2s}\!({\bar{\gamma}}_{g}^{j}(q))
^{-1} \sum_{r=0}^{2s-h}[^{2s-h}_{\;\;\;r}]_{-q^{-1}}(-1)^{r
p_{j_1,h'_1}}q^{r(h'_1)}\prod_{r_1=h'_1}^{h_1+1}\gamma_{r_1}^
{j_1}(q)\prod_{r_2=h'_2}^{h_2+1}\gamma_{r_2}^{j_2}(q)
\;C'_{\mathcal{I}}\!(^{j_1,j_2,s}_{h'_1,h'_2,2s}),\label{cij} \eea
up to the additive term like
$C_{\mathcal{I}}(^{j_1,j_2,j}_{h_1,h_2,h})P(\gamma^s_i,\bar{\gamma}^s_{i'},
\tilde{\gamma}^s_{i''})$, where $P(\gamma^s_i,\bar{\gamma}^s_{i'},
\tilde{\gamma}^s_{i''})$ is a rational function.
\section{Even dimensional (unconventional) representations and \\ the quantum
algebras $osp_q(1|2)$ and $s\ell_t(2)$  at roots of unity }

To complete our analysis we observe also the emerging of the
indecomposable representations in the fusions for the even
dimensional irreducible representations, which have no classical
counterparts (see the second section). Let us denote $V_r=\{
(1-r)/2+(\imath\pi/(2\log{q})),(3-r)/2+(\imath\pi/ (2\log{q}))
,...,(r-1)/2+(\imath\pi/(2\log{q})) \}$  by $V_{4j+1}$, with
$2j=\frac{r-1}{2}$.

In accordance with (\ref{t}), for the general values of $q$,
decomposition of the tensor products for the representations with
odd and even dimensions has form (\ref{v*v}), but now with $j$ is
taking integer, half-integer or quarter integer values.

\paragraph{Representation $V_r$, $r\in 2\mathbb{Z}_{+}$.} {Let} us present the action of the algebra on the vectors
$\{v_{j}(h)\},\quad
h=\!-\!2j+\!\lambda,-\!2j+\!1+\!\lambda,...,2j+\!\lambda$, of the
\textit{even dimensional irreducible representation} $V_{4j+1}$
($j$ is a quarter integer) as follows
 \bea 
&&\left\{\ba{ll}k\cdot v_j(h)=q^{h} v_j(h),\quad &\\
 e \cdot
v_j(h)=v_j(h+1),\quad &\quad e \cdot v_j(2j+\lambda)=0,\\
 f\cdot v_j(h)=\alpha^{j}_{h} (q) v_j(h-1),\quad & \quad f \cdot
v_j(-2j+\lambda)=0,\ea\right.\qquad\qquad\qquad\qquad\qquad
\\\label{alpha1} &&\alpha^{j}_{h}
(q)=\sum_{i=0}^{2j-h+\lambda}(-1)^{2j-(h-\lambda)-i}[2j+\lambda-i]_q,\qquad
h=-2j+1+\lambda,...,2j+\lambda.
 \eea
Here in comparison with the previous cases (\ref{odd}, \ref{ind}),
we have specified parameters $\beta$ and $\gamma$.

Recall that irreducibility of $V_{4j+1}$ turns to be spoiled, if
at least one of the functions $\alpha_{h}^{j}(q)$ vanishes,
indicating the existence of a proper sub-representation.

In the case under consideration (\ref{alpha1}), taking into
account that $q^{2\lambda}=-1$, the equation $\alpha_{h}^{j}(q)=0$
is equivalent to (below the notation $\bar{h}=h-\lambda$ is used)
 \be
(1-(-1)^{2j+\bar{h}}q^{2j+\bar{h}})(1+(-1)^{2j-\bar{h}
}q^{2j+1-\bar{h}})=0. \label{hq}
 \ee
It follows from the property $\alpha^{j}_{r+\lambda}(q)=-\alpha
^{j}_{-r+\lambda+1}(q)$, that two multipliers in the l.h.s. of
(\ref{hq}) are equivalent. Hence one can consider only one of
them, say $(1-(-1)^{2j+\bar{h}}q^{2j+\bar{h}})=0$, which has
solution of the form:
 \be \mbox{If}\quad q^{2n}=1, \quad n\in \mathbb{N},
\quad\mbox{then} \qquad \bar{h}=2n p-2j, \;\;p\in \mathbb{N},\quad
0<2n p\leq 4j. \label{qn1}\ee
 \be
\mbox{If}\quad q^{2n+1}=-1, \quad n\in \mathbb{N},\quad
\mbox{then}\qquad \bar{h}=(2n+1) p-2j, \;\;p\in \mathbb{N}\quad
0<(2n+1)p\leq 4j.\label{qn2}
 \ee
If $j_1$ is the weight $\bar{h}/2$ corresponding to $\alpha_{h+1}^
{j}(q)=0$, for which $|h-\lambda|$ takes its maximal value, then
$\bar{V}_{4j+1}$ has proper sub-representation $V_{4j_1+1}$. And
according to the above observations (see Statement II), in the
fusions such representations $\widehat{\bar{V}_{4j+1}\oplus
V_{4j_1+1}}= \mathcal{I}^{4(j+j_1)+2}_{\{4j+1,4j_1+1\}}$
 can appear. The values of $j_{max}$ 
 can be defined from the formulas (\ref{jmax1},
 \ref{jmax2}) adding the factor $\frac{1}{4}$.


It follows from (\ref{qn1}, \ref{qn2}) that the dimensions of
$\mathcal{I}$-representations are again defined by the formula
(\ref{in}). Their structures are described in (\ref{ind}). An
example is the indecomposable representation at $q^3=-1$,
$\mathcal{I}^{(6)}_{\{4,2\}}=\widehat{\bar{V_4}\oplus V_2}$,
studied in detail in the third subsection (\ref{23}). So for the
given $n,\; q^n=\pm 1$, inclusion of the even dimensional irreps
could enlarge the class of the representations $\mathcal{I}^{2n
p}$ with representations $\mathcal{I}^{(2n p)}_{\{2r,2n p-2r\}}$,
which can appear in the mixed fusions as $V_{2s_1}\otimes V_{2
s_2+1}$.

 \paragraph{Note.} {\it And there is an interesting fact}, which we would like to
mention. As it was stated the odd dimensional irreps of
$osp_q(1|2)$ form a closed fusion at general $q$. However for the
cases, when $q$ is a root of unity and $\lambda=\imath\pi/
(2\log{q})$ is a rational quantity, in the decomposition of the
multiple tensor products of the conventional irreps such
indecomposable representations can arise, which have even
dimensional irreps' origin: in (\ref{rp}) $\mathbf{r_p}$ accepts
an even number, when $\mathcal{N}$ is odd and $p$ is even. In the
mentioned cases in $\otimes^n V_3$ a direct sum $V_r\oplus
V_{p\mathcal{R}-r}$ at $(-q)^{\mathcal{R}/2}=1$ ($r>r_{max}$)
replaces with $p(
\widehat{\bar{V}_{\mathcal{R}-\mathbf{r}_p}\oplus
V_{\mathbf{r}_p}})$, and while $r,\; p\mathcal{R}-r$ are odd
numbers, the dimensions $\mathbf{r}_p$ and
$\mathcal{R}-\mathbf{r}_p$ are even.

\paragraph
{ { The resemblance of the representations of the algebras
$osp_q(1|2)$ and $s\ell_t(2)$  at roots of unity. }}
 The correspondence between the odd
dimensional conventional irreps of the quantum super-algebra
$osp_q(1|2)$ and the odd dimensional non-spinorial irreps of the
quantum algebra  $s\ell_t(2)$ ($t=i q^{1/2}$) was mentioned and
investigated in the works \cite{S, z}. The consideration of the
even dimensional irreps of $osp_q(1|2)$ \cite{Ki} made the
equivalence of the finite dimensional irreducible representations
of $osp_q(1|2)$ and $s\ell_t(2)$ at general $q$ complete. About
the correspondence of the $R$-matrices and Lax operators
with symmetry of $osp_q(1|2)$ and $s\ell_t(2)$ there are
discussions in the works \cite{S,Ki,SHKKH}.

 The classification and
investigation of the finite dimensional representations of
$s\ell_t(2)$, when $t$ is a root of unity can be found in the
works \cite{PS, Ar1}. For any $\mathcal{N}_t$
[$\mathcal{N}_t=\{^{N/2,\;even\;N}_{N,\;\;odd\;\;N}\}$, $t^N=1$],
the lowest weight indecomposable representation, emerging from the
fusions of the irreps, has dimension $2\mathcal{N}_t$.

The relation $t^{2N}=(-1)^N q^{N}$ helps us to connect the
dimensions of the indecomposable representation
$\mathcal{I}^{\mathcal{R}}$ of the algebra $osp_q(1|2)$ with the
respective ones of $s\ell_t(2)$. The first possibility in
(\ref{inj}) corresponds to the relation $t^{2N}=-1$, i.e.
$\mathcal{N}_t=2N$, the second one corresponds to $t^{2N}=1$, i.e.
$\mathcal{N}_t=N$. This means, that the dimension of the
representation $\mathcal{I}^{\mathcal{R}}$  can also be presented
as $2\mathcal{N}_t$. In the same way the correspondence of the
dimensions of the permissible irreps can be stated.

 As we  see, the mentioned equivalence of the
representations of two quantum algebras can be extended also for
the exceptional values of $q$ \cite{S}. All the tools and
principles which are used in this paper (Clebsh-Gordan
decomposition, Statements and et al.) are valid also in the case
of the algebra $s\ell_t(2)$. And this similarity can help us to
compare the analysis of the fusion rules when $q$ is a root of
unity with the known schematic results \cite{Ar1, KKH} and to be
convinced of their correspondence, and also to extend the detailed
analysis of the fusion rules of the multiple tensor products of
the irreps and indecomposable representations to the case of
$s\ell_t(2)$.

\section{Summary}

We studied the lowest weight representations of $osp_q(1|2)$ at
{\textit{the exceptional values of $q$} (when $q$ is a root of
unity)}, and as result we listed all the possible irreps and
indecomposable representations appearing for given $N,\; q^N=\pm
1$, and formulated the modification of the conventional fusion
rules. We described how and when indecomposable representations
appear in the decompositions of the tensor products. It led to a
scheme for explicit construction of the decompositions for the
tensor products of both irreducible and indecomposable
representations when deformation parameter takes exceptional
values.

\section*{\large Acknowledgments}
 The work was partially supported by the Volkswagen Foundation of
Germany and INTAS grant No 03-51-5460. D.K. thanks High Energy
section of ICTP for kind hospitality.

\section{Appendix}
\subsection*{The tensor product decompositions by direct constructions}
\setcounter{section}{0}\setcounter{footnote}{0}
\addtocounter{section}{0}\setcounter{equation}{0}
\renewcommand{\theequation}{A.\arabic{equation}}
\paragraph{The eqs. (\ref{vqnn}, \ref{vqn}).}
Here we would like to return once again to the formulas
(\ref{vqnn}, \ref{vqn}).
  Due to associativity of the co-product,
the representation of type $\mathcal{I}$ can arise only from the
degeneration of the decomposition of the following tensor product
(below we use notation $j_{max}\equiv J$ 
)
 \be
V_{4J+1}\otimes V_{3}=V_{4J-1}\oplus V_{4J+1}\oplus V_{4J+3}.
\label{vj}
 \ee
Let us denote the vector states of the irreps in the l.h.s. of
(\ref{vj}) by $u_{J}(h)$ and $u_{\frac{1}{2}} (h)$. After
calculations of the corresponding coefficients (\ref{cgc}) and
inserting them in (\ref{vjh}), one obtains the following
expressions (up to the common multipliers) for the  vector states
$v_{j}(2J)$ of the representations in the r.h.s of equation
(\ref{vj})
 \bea
\!v_{J+\frac{1}{2}}(2J) &=&
[2J]_q \; \left(u_{J}(2J\!-\!1)\otimes
u_{\frac{1}{2}}(1)\right)+q^{2J}\left( u_{J}(2J)\otimes
u_{\frac{1}{2}}
(0)\right),\\
\!\!\!\!\!\!\!\!\!\!v_{J}(2J) &=& \left(u_{J}(2J\!-\!1)\otimes
u_{\frac{1}{2}}(1)\right)-q^{-1}\; \left(u_{J}(2J)\otimes
 u_{\frac{1}{2}}(0)\right).
 \eea
So, recalling (\ref{jmax1}), one sees that for the cases $q^N=1$,
with odd integer $N$, or $q^N=-1$, with even integer $N$, the
relation $v_{J +\frac{1}{2}}(2J)=(-1)^N v_{J}(2J)$ takes place.
The relations among the remaining vectors $v_{j}(h)$ can be
obtained by the repeated actions of the lowering operator $f$. As
the action of the operators $f^n$, $n < 4J$, on the states
$v_{J\!+\!\frac{1}{2}}(2J)$ and $v_{J}(2J)$ does not annihilate
them ( $V_{4J+1}$ is an irreducible representation), we can take
$\gamma^{J\!+\!\frac{1}{2}}_h(q)=1,\;\gamma^{J}_h(q)=1, h=\{-2
J+1,...,2J\}$. Then
 \bea
\left\{v_{J\!+\!\frac{1}{2}}(2J),
 ...,
 v_{J\!+\!\frac{1}{2}}(-2J)\right\}=(-1)^N
\left\{v_{J}(2J),...,
 v_{J}(-2J)\right\}.\label{vu1}\eea
In the same way we check, that the vectors
$v_{J+\frac{1}{2}}(2J\!-\!1)$ and $v_{J-\frac{1}{2}}(2J\!-\!1)$
are expressed by these formulas respectively
  \bea\nn
&\!\!\!\!\!\!\!\!\!\! 
[2J]_q
([2J]_q\!-\![2J\!-\!1]_q)u_{J}(2J\!-\!2)\otimes u_{\frac
{1}{2}}(1)+q^{4J}u_{J} (2J)\otimes
u_{\frac{1}{2}}(-1)
-q^{2J}(1\!-\!\frac{1}{q})[2J]_q
 u_{J}(2J\!-\!1)\otimes
 u_{\frac{1}{2}}(0)&
 \\\!
&\!\!\!\!\!\!\!\!\!\!\mbox{and}\quad 
u_{J}(2J\!-\!2)\otimes
u_{ \frac{1}{2}}(1)\!-\! q^{-1} u_{J}(2J)\otimes
u_{\frac{1}{2}}(-1)+q^{-1} u_{J}(2J\!-\!1)\otimes
u_{\frac{1}{2}}(0),& 
 \eea
and we make sure that two vectors become linearly dependent, for
$q^N\!=\!-1, \; N\!=4J\!+\!1$ is odd 
(\ref{jmax2}): 
%
 \bea \{v_{J\!+\!\frac{1}{2}}(2J\!-\!1),
 ...,v_{J\!+\!\frac{1}{2}}(\!-\!2J\!+\!1)\}=
 \{v_{J\!-\!\frac{1}{2}}(2J\!-\!1),...,
 v_{J\!-\!\frac{1}{2}}(\!-\!2J\!+\!1)\}.\label{vu2}\eea
One can construct representations $\upsilon'$, demanding $e\cdot
v'(2J)= v_{j_2} (2J+1)$ for the first case (\ref{jmax1},
\ref{vu1}) and $e\cdot v'(2J-1)= v_{j_2}(2J)$ for the second case
(\ref{jmax2}, \ref{vu2}). The solutions to these equations are not
unique (the solutions to the homogeneous equations, i.e.
$v_{J+\frac12} (2J)$ and $v_{J+\frac12} (2J-1)$, can be added).
The remaining vectors of $\mathcal{I}$ it is possible to construct
by the action of the lowering generator $f$ on the vectors
$v'(2J)$ or $v'(2J-1)$. In the case (\ref{jmax1}) the solution can
be taken in the following form
 \bea \label{vs1}&v'(2J)=
u_{J}(2J \!-\!1)\otimes u_{\frac{1}{2}} (1),
\quad v'(h')=f^{2J-h} v'(2J),\;\; h'=2J-1,...,-2J.&\eea The
resulting representation $\{v_{J+1/2}(h),v'(h')\}$, with
$h=2J+1,...,-2J-1,\;h'= 2J,...,-2J$, consists with the
indecomposable representation $\mathcal{I}^{(8J+4)}_{\{4J+3,4j
_{max}+1\}}$. This is in the agreement with the formula
(\ref{vqnn}), as $J=(N-1)/2$.

In the second case (\ref{jmax2}, \ref{vu2}), one can check that
the vectors
  $\{v_{J+1/2}(h),v'(h')\}$,  $h=2J+1,...,-2J-1,\;h'=
  2J-1,...,-2J+1$, are forming indecomposable representation
  $\mathcal{I}^{(8J+2)}_
  {\{4J+3,4J-1\}}$ (see (\ref{vqn})), with the following vectors $v'(h')$
\bea\label{vs2} & v'(2J-1)= q \;u_{J}(2J\!-\!2)\otimes
 u_{\frac{1}{2}}(1)+u_{J}(2J)\otimes
 u_{\frac{1}{2}}(-1),
\quad v'(h')=f^{2J-h} v'(2J-1).& \eea
Here $
  h'=2J-2,...,-2J+1$. By using the algebra relations, it is easy to check that
representations $\mathcal{I}^{(8J+4)}_{\{4J+3,4J+1\}}$ and
 $\mathcal{I}^{(8J+2)}_{\{4J+3,
 4J-1\}}$  are satisfying  (\ref{ind}), and to obtain the
coefficients
$\beta,\gamma,\bar{\beta},\bar{\gamma},\tilde{\beta},\tilde{\gamma}$.
\paragraph{The case of (\ref{ip}).}

The representations $\mathcal{I}^{(p\mathcal{R})}$, $p>1$, appear
in the tensor product decompositions $\otimes^n V_3$ quite
similarly as $\mathcal{I}^{(\mathcal{R})}$, but now due to the
deformation of a sum $V_{4j_1+1}\oplus V_{4j_2+1}$ in the
decomposition, with $j_2>j_1>J$ (as $4j_2+4 j_1+2=p\mathcal{R}$,
see (\ref{inj})). It can be checked analogously to the previous
case, using Clebsh-Gordan coefficients, taking now tensor product
$\bar{V}_{4j+1}\otimes V_3$  ($j>J$).
 The difference here is
that
 there are another highest and lowest weights also in
the representation $\bar{V}_{4j_2+1}$ besides of the weights $\pm
j_{1(2)}$, as $\Delta(f^{\mathcal{N}})$ (\ref{fN}) (as well as the
operator $\Delta(e^{\mathcal{N}})$) vanishes when
$(-q)^{\mathcal{N}}=1$, and now ${\mathcal{N}}\leq 4 j_1$. In the
paragraph after (\ref{rp}) we have denoted them as $j_2^{\; i}$,
$i=1,...,p=[(4j_2+1)/\mathcal{N}]$.

In the same way, as above, at the values of $q$ defined by
(\ref{in}) a linear dependence is established between the 
vectors $f^{n}v_{j_2}(2j_2^{\; i})$, $0\leq n<\mathbf{r_p}$,
$i=1,..,p$ (note, that $j_2^{\; p}=j_1$),
of $\bar{V}_{4j_2+1}$ and the corresponding vectors of
$\bar{V}_{4j_1+1}$ with the same $h$.
Solving the equations $e v'(2j_2^{\; i})=v_{j_2}(2j_2^{\; i}+1)$,
one constructs all $v'$-vectors, $v'(2j_2^{\; i}-n)=f^n
v'(2j_2^{\; i})$, $0\leq n<\mathbf{r_p}$.

\end{document}